\documentclass[aps,pra,showpacs,twocolumn,amsmath,a4paper]{revtex4}
\usepackage{geometry}
\usepackage{bbm}
\usepackage{amssymb}
\usepackage{amsmath}
\usepackage{amsfonts}
\usepackage{epsfig}
\usepackage{amsthm}
\usepackage{mathdots}

\usepackage{dsfont}

\usepackage{verbatim}

\newcommand{\bra}[1]{\ensuremath{\left\langle #1\right|}}
\newcommand{\ket}[1]{\ensuremath{\left|#1\right\rangle}}
\newcommand{\braket}[2]{\ensuremath{\left\langle #1\vphantom{#2}\right.\left|\vphantom{#1}#2\right\rangle}}

\newcommand{\one}{\mbox{$1 \hspace{-1.0mm}  {\bf l}$}}

\def\C{\hbox{$\mit I$\kern-.7em$\mit C$}}
\def\R{\hbox{$\mit I$\kern-.6em$\mit R$}}
\def\tr{\mathrm{tr}}
\newcommand{\F}{\mathbb{F}}

\newcommand{\bea}{ \begin{eqnarray} }
\newcommand{\eea}{ \end{eqnarray} }
\def\bi{\begin{itemize}}
\def\ei{\end{itemize}}
\def\bc{\begin{center}}
\def\ec{\end{center}}

\newtheorem{theorem}{Theorem}
\newtheorem{corollary}[theorem]{Corollary}
\newtheorem{lemma}[theorem]{Lemma}

\begin{document}

\title{Graph--state formalism for mutually unbiased bases}

\author{Christoph Spengler}
\email{Christoph.Spengler@uibk.ac.at}
\author{Barbara Kraus}
\affiliation{Institute for Theoretical Physics, University of Innsbruck, Innsbruck, Austria}

\begin{abstract}
A pair of orthonormal bases is called mutually unbiased if all mutual overlaps between any element of one basis with an arbitrary element of the other basis coincide. In case the dimension, $d$, of the considered Hilbert space is a power of a prime number, complete sets of $d+1$ mutually unbiased bases (MUBs) exist. Here, we present a novel method based on the graph--state formalism to construct such sets of MUBs. We show that for $n$ $p$--level systems, with $p$ being prime, one particular graph suffices to easily construct a set of $p^n+1$ MUBs. In fact, we show that a single $n$--dimensional vector, which is associated with this graph, can be used to generate a complete set of MUBs and demonstrate that this vector can be easily determined. Finally, we discuss some advantages of our formalism regarding the analysis of entanglement structures in MUBs, as well as experimental realizations.
\end{abstract}


\maketitle

\section{Introduction}
A density matrix of a $d$--level quantum system is described by $d^2-1$ real parameters. Since a von Neumann measurement can reveal at most $d-1$ independent probabilities, $d+1$ such measurements are at least necessary to determine the state of the system. The question whether, in certain cases, $d+1$ measurements are also sufficient led to the introduction of mutually unbiased bases \cite{Schwinger,Iv,Durtreview}. A pair of orthonormal bases, say $\mathcal{B}_{k}=\{|i_{k}\rangle \}_{i=0}^{d-1}$ and $\mathcal{B}_{l}=\{|j_{l}\rangle \}_{j=0}^{d-1}$, of a $d$--dimensional Hilbert space, $\mathcal{H}=\mathbb{C}^{d}$, is called mutually unbiased if $|\langle i_{a}|j_{b}\rangle |^{2}=\frac{1}{d}$ holds for any choice of elements $i$ and $j$. If for a set of bases $\{\mathcal{B}_{k}\}$ this relation holds true for all possible pairs of bases, i.e. $|\langle i_{k}|j_{l}\rangle |^{2}=\frac{1}{d}$ for all $i,j$ and $k\neq l$, this set is called a set of \emph{mutually unbiased bases} (MUBs). A simple example of a set of three MUBs for dimension $d=2$ are the normalized eigenvectors of the three Pauli operators.

As discussed in Refs.~\cite{wootters,Adamson}, MUBs show that state tomography with the minimum number of $d+1$ measurements is indeed possible. In fact, such bases maximize the information extraction per measurement and minimize the effects of statistical errors \cite{wootters}. Besides state tomography, MUBs play an important role in quantum key distribution \cite{Brusssecret,Cerfsecret}, and solutions to the so-called mean king problem \cite{meanking1,meanking2}. Moreover, they have recently been shown to be useful for entanglement detection \cite{mubdetection}. Furthermore, it was discovered that they have interesting connections to symmetric informationally complete positive-operator-valued measures \cite{wootterspovms}, and complex $t$--designs \cite{MUBdesigns,Grossdesign}.

The reason why MUBs have found several of the applications mentioned above is mainly due to the fact that if a system is prepared in one of the states constituting a particular basis $\mathcal{B}_k$, then any measurement outcome of an observable whose eigenbasis, $\mathcal{B}_l$, is mutually unbiased with respect to $\mathcal{B}_k$ is equally likely. Consequently, pairs of observables whose eigenbases are mutually unbiased are complementary. In particular, there is no state such that the outcome with respect to both observables is predictable with certainty; a fact which is exploited, for instance, in quantum key distribution protocols.

Whereas $d+1$ MUBs are sufficient to reconstruct a density matrix of a $d$--dimensional system, it is \emph{a priori} not clear how many such bases exist for a given dimension $d$. However, it is easy to show that $d+1$ is not only the required number of MUBs for complete state tomography, but also the maximum number of MUBs \cite{wootters}. For this reason, $d+1$ MUBs are called a \emph{complete set} of MUBs. For any dimension $d$ which is a power of a prime number, it was shown via an explicit construction that there always exists a complete set of MUBs \cite{wootters}. However, for all remaining dimensions not even a single example of a complete set is known. In fact, there is evidence that, in general, there exists no such complete set. For instance, recent numerical searches in Refs.~\cite{Butterley,Brierley1,Raynal}, and analytical investigations in Refs.~\cite{Brierley2,Brierley3,jaming09,BruknerLat,Bengtssond6}, indicate that there are only three MUBs in dimension $d=6$, whereas a complete set would consist of seven. Nevertheless, a rigorous proof for the non--existence of complete sets of MUBs in non--prime power dimensions is still missing.

The original construction of complete sets of MUBs for Hilbert spaces of dimension $d$ with $d$ being an odd prime number is based on quadratic exponential sums \cite{Iv}. This method was later generalized to $d$ being a power of a prime number by making use of the theory of finite extension fields, or Galois fields \cite{wootters}. It is based on so-called Weil sums \cite{KlappenRott}. A different method to construct complete sets of MUBs for dimensions of prime powers, was presented in Ref.~\cite{Bandyopadhyay}. Herein, it was shown that MUBs can be extracted from a partition of the associated operator space into certain commuting sets. These methods have then also been used to construct so-called unextendible MUBs~\cite{Unextendibles}. As will become clearer later, both methods have their advantages compared to the other. Whereas the first one can be easily used to generate MUBs, once the group theoretic results are applied, the second is solely using the properties of generalized Pauli operators. However, in order to construct the desired complete set, some relations between these operators have to be verified.

In this paper, we present an alternative formalism to construct complete sets of MUBs. The key idea is to use tools form quantum information theory, rather than abstract mathematical concepts. Here, our starting point is the fact that, with respect to the computational basis, the elements of corresponding mutually unbiased bases belong to the class of so-called \emph{locally maximally entanglable} (LME) states \cite{LME}. Those states can be generated solely by applying phase gates to an initial state, which is the equally-weighted superposition of all computational basis states. A special class of LME states are the so-called \emph{graph--states} \cite{Hein,labeledqubits,quditgraphs1,quditgraphs2}, where operations on the initial state are restricted to $2$--body interactions with a particular fixed phase. These states play a key role in a variety of quantum information processing schemes such as for example quantum error correction (see e.g. Ref.~\cite{quditgraphs1,quditgraphs2} and references therein), and measurement-based quantum computing \cite{mbqc}. Here, we show that a minor generalization of graph--states, where $1$--body phase gates are also allowed, may be utilized to construct MUBs. As in the two previously mentioned constructions, we obtain a simple sufficient condition for mutual unbiasedness on the adjacency matrices defining the generalized graph--states. Then, we show how this condition can be met for a complete set of MUBs using symmetric matrix representations of finite fields. We show that such representations always exist for all prime power dimensions, and give a constructive algorithm for obtaining them. This concept allows us to prove that a single symmetric matrix whose characteristic polynomial cannot be factorized (i.e. is irreducible) is sufficient to construct a complete set of MUBs for any prime power dimension. Furthermore, for multipartite qubit systems, we show that a complete set of MUBs may be encoded in a single $n$--dimensional vector whose components are the diagonal elements of a tridiagonal matrix. Here, we show that one can either consider the generalized graph--states corresponding to all $p^n-1$ powers of those matrices or the ones corresponding to arbitrary linear combinations of the first $n$ powers of them. In the first case, we call the corresponding graph--state a \emph{primitive graph--state}, whereas the set of graph--states occurring in the latter case are called \emph{fundamental graph--states}. As we will see, the graphical representation of generalized graph--states will make it possible to easily extend the $n$ fundamental graph--states to a set of states corresponding to a complete set of MUBs. These simple and constructive methods lead to a set of $p^n+1$ MUBs for all prime power dimensions.

Our results also provide a general method for implementing complementary measurements by means of quantum circuits consisting of a few elementary gates. First attempts in this direction have recently been made in Ref.~\cite{Seyfarth} for a restricted class of qubit systems, and in Ref.~\cite{Wiesniak} for bipartite systems of prime dimension. Here, we present a complete framework for constructing MUBs using only two local operations and one entangling gate. In particular, our scheme does not only work for the special cases discussed in Refs.~\cite{Seyfarth,Wiesniak}, but for \emph{all} multipartite prime-dimensional systems. Besides the fact that our formalism is mathematically simple, it also allows to easily address questions related to the presence of entanglement in basis states of complete sets of MUBs. Questions of this type have been considered in Refs.~\cite{Wiesniak,LawrenceEntPatterns,Romero}; however, not much is known about the entanglement structure in MUBs beyond tripartite systems. In this respect, our descriptive formalism in terms of graphs may lead to new insights on the role of entanglement in MUBs for more complex many-body systems.

The remainder of the paper is organized as follows. In Section \ref{SECConstr}, we briefly review the concept of finite fields and their extension, as well as the two most commonly used constructions of MUBs. In Section \ref{SECGraph}, the generalized graph--state formalism for multipartite-multilevel systems is introduced. Subsequently, for a pair of bases whose elements are generalized graph--states, we derive a sufficient condition for mutual unbiasedness in Section \ref{SECMUBGraph}. In Section \ref{SECCompMUBs}, we present a simple method to construct complete sets of MUBs in terms of generalized graph--states for all prime power dimensions. As mentioned before, in contrast to other construction, we derive a very simple construction, which is based on a single graph. We demonstrate that the complete set of MUBs can then be easily obtained from this graph. Moreover, we show that the MUBs can be easily read off from the graphical representation of $n$ fundamental graphs. In Section \ref{SECentstrut}, we use the graph--state formalism to study aspects of entanglement for MUBs in some examples. A connection between the adjacency matrices corresponding to complete sets of MUBs and the average purity of reduced density matrices is established in Section \ref{SECDesign}. Finally, we discuss the experimental implementation of complementary measurements using a sequence of one-body and two-body phase gates in Section \ref{SECImplement}, and give a brief conclusion in Section \ref{SECConclusion}.

\section{Preliminaries and established constructions} \label{SECConstr}
In this section, we first give a brief summary of some basic concepts related to finite fields and their extensions and then summarize two most commonly used constructions of MUBs for prime power dimensions introduced by Wootters and Fields in Ref.~\cite{wootters}, and Bandyopadhyay \emph{et al.} in Ref.~\cite{Bandyopadhyay}. Readers who are already familiar with these constructions might just want to have a brief glance at this section in order to become familiar with the notation used throughout the paper.

\subsection{Finite fields and their extensions} \label{IntroFF}

A finite field, $\mathbb{F}_d=(S_d,+, \cdot )$, is defined as a set $S_d$, with finitely many elements $|S_d|=d$, on which two binary operations $+$ (addition) and $\cdot$ (multiplication) are defined such that $(S_d,+)$ and $(S_d \backslash \{0\}, \cdot)$ form abelian groups, and $\alpha \cdot(\beta+\gamma)=\alpha \cdot \beta + \alpha \cdot \gamma$ for all $\alpha,\beta,\gamma \in S_d$. Here, the element $0$ denotes the neutral element of the additive group $(S_d,+)$), and $1$ the neutral element of the multiplicative group $(S_d \backslash \{0\}, \cdot)$. Furthermore, $-\alpha$ represents the additive inverse of $\alpha \in S_d$, i.e. $\alpha + (-\alpha) = \alpha - \alpha =0$, and $\beta^{-1}$ denotes the multiplicative inverse of $\beta\in S_d\backslash \{0\}$, i.e. $\beta \cdot \beta^{-1}=1$. Finite fields were shown to exist iff the number of elements of $S_d$ is a prime power, i.e. $d=p^n$ where here and in the following $p$ is a prime number, and $n$ is an arbitrary integer \cite{finitefields,Lidl}.

Prime fields $\mathbb{F}_p$ are isomorphic to $\mathbb{Z}_p=(\mbox{$\{0,\ldots,p-1\}$}, +, \cdot )$, i.e. the set of integers $\{0,\ldots,p-1\}$ with addition ($+$) and multiplication ($\cdot$) performed modulo $p$. An extension of a field, $\F_d$, is a field (under the operations of $\F_d$) which contains $\F_d$. A prime field $\mathbb{F}_p \cong \mathbb{Z}_p$ can be extended to a prime power field $\mathbb{F}_{p^n}$ for an arbitrary integer $n$ as follows. Consider a monic polynomial \footnote{A polynomial is called monic if the leading coefficient is 1.}, $f(x)=x^n+c_{n-1}x^{n-1}+\ldots+c_1 x +c_0$, of degree $n$ with coefficients $c_i \in \mathbb{Z}_p$ which cannot be factorized over $\mathbb{Z}_p$, i.e. a so-called \emph{irreducible} polynomial. A necessary condition for irreducibility is that the polynomial does not have a root in $\mathbb{Z}_p$. Let us denote by $\alpha \notin \mathbb{Z}_p$ one of the roots of $f(x)$, i.e. $f(\alpha)=0$. The elements of an extension field $\F_{p^n}$ can then be represented by all polynomials in $\alpha$ over $\mathbb{Z}_p$ up to degree $n-1$, i.e. $\{1,\alpha,\alpha^2,\ldots,\alpha^{n-1} \}$ is a basis of $\F_{p^n}$. The multiplication ($\cdot$) and addition ($+$) of these elements is performed modulo $f(\alpha)$. In other words, the extension field $\mathbb{F}_{p^n}$ can be viewed as the residue class ring of $\mathbb{F}_p[x]/(f(x))$, i.e. the ring of polynomials with coefficients in $\mathbb{Z}_p$ modulo the irreducible polynomial $f(x)$ \footnote{Note that this is like the extension of the real numbers $\mathbb{R}$ to the complex numbers $\mathbb{C}$, where the symbol $\alpha=\mathbbm{i} \notin \mathbb{R}$ represents the root of $f(x)=x^2+1$.}. Since
each of those polynomials can be written as a linear combination of the basis elements (and thus are characterized by $n$ coefficients $c_i \in \mathbb{Z}_p$), they can be represented by an $n$--dimensional vector $(c_0,\ldots,c_{n-1})$. The number of different polynomials, i.e. the number of elements of $\mathbb{F}_{p^n}$, is therefore $p^n$. It is important to note that neither the choice of the irreducible polynomial nor the choice of the root changes the structure of the extension field, in the sense that they are all isomorphic. As an example we consider the prime field $\mathbb{Z}_3$ and the irreducible polynomial $f(x)=x^2+x+2 \in \mathbb{Z}_3[x]$. Let us denote by $\alpha$ one of the roots of $f(x)$. The extension field $\F_{3^2 }$ then consists of the following nine elements, $0,1,2,\alpha, \alpha+1,\alpha+2,2\alpha,2\alpha+1,2\alpha+2$, i.e. all polynomials over $\mathbb{Z}_3$ with degree smaller than two.

The \emph{minimal polynomial} of an element $\gamma \in \mathbb{F}_{p^n}$ is defined as the monic polynomial $p(x)=x^m+ c_{m-1}x^{m-1} + \ldots + c_{1}x +c_0$ over $\mathbb{Z}_p$ of smallest degree $m$ for which $p(\gamma)=0$. Every element of $\mathbb{F}_{p^n}$ has a unique minimal polynomial, which is necessarily irreducible (over $\mathbb{Z}_p$). If the minimal polynomial of an element $\gamma$ is of the order $n$, then the set of its powers $\{\gamma^i\}_{i=0}^{n-1}$ constitutes a basis of $\mathbb{F}_{p^n}$, i.e. every element of $\mathbb{F}_{p^n}$ can be uniquely represented as $b_0+b_1 \gamma + \ldots + b_{n-1} \gamma^{n-1}$.

Since the multiplicative group $\mathbb{F}_{p^n}\backslash\{0\}$ is cyclic \cite{Lidl} (as the multiplicative group of any finite field), it contains a so-called \emph{primitive element} $\gamma$ with the property that its first $p^n-1$ powers generate all non--zero elements of the field, i.e. $\F_{p^n} \backslash \{0\}=\{\gamma^i \}_{i=0}^{p^n-2}$. The minimal polynomial of a primitive element is called a \emph{primitive polynomial}. The crucial characteristic of a primitive polynomial $p(x)$ is that the smallest positive integer $m$ for which it becomes a factor of the polynomial $x^m-1$ over $\mathbb{Z}_p$ is $m=p^n-1$. That is, if $p(x)$ is a primitive polynomial then for any $m < p^n-1$ there exist no polynomial $g(x)$ over $\mathbb{Z}_p$ such that $x^m-1=g(x)p(x)$. Every such polynomial is necessarily of order $n$ and irreducible. Hence, if a primitive polynomial is used from the beginning as the irreducible polynomial $f(x)$ for which $f(\alpha)=0$, to construct the extension field $\mathbb{F}_{p^n}$, then the element $\alpha$ is itself a primitive element and therefore $\F_{p^n} \backslash \{0\}=\{\alpha^i \}_{i=0}^{p^n-2}$. A list of irreducible and primitive polynomials can be found in Ref.~\cite{Lidl,primpolylist}. In addition, nowadays there are also several commercial software packages, such as the `Communications System Toolbox' for Matlab$\textsuperscript{\textregistered}$, which are able to automatically generate those polynomials.

Finally, let us note that for an element $\gamma \in \mathbb{F}_{p^n}$, the \emph{trace operator} is defined as
\bea
\label{traceop} \tr(\gamma)=\sum_{k=0}^{n-1} \gamma^{p^k}=\gamma+\gamma^p+ \ldots +\gamma^{p^{n-1}}.
\eea
It can be shown that the trace operator is a linear map from $\mathbb{F}_{p^n}$ to $\mathbb{F}_p \cong \mathbb{Z}_{p}$.

\subsection{MUBs from finite field extensions } \label{woottconstr}

An important result within field theory, which was used to construct complete set of MUBs, is that
\begin{align}
\label{charsum} \left| \sum_{l\in \F_{p^n}} \omega_p^{\tr(kl^2+ml)} \right|=\sqrt{p^n} \ ,
\end{align}
for $p\geq 3$ prime, arbitrary $m \in \mathbb{F}_{p^n}$, non--zero $k \in \mathbb{F}_{p^n}$ and $\omega_p=e^{2\pi \mathbbm{i} /p}$. Using this relation, it follows immediately that the following set constitutes a complete set of MUBs for all odd prime power dimensions \cite{wootters,Iv}. The first basis is the computational basis $\mathcal{B}_C$. The remaining $d$ bases $\mathcal{B}_{k}=\{\ket{v_k(m)}\}_{m\in \mathbb{F}_{p^n}}$, which are pairwise mutually unbiased, are given by \begin{align}
\label{oddcase}
\ket{v_k(m)}=\frac{1}{\sqrt{d}} \sum_{l\in \mathbb{F}_{p^n} } \omega_p^{\tr(kl^2+ml)}\ket{e(l)} \ ,
\end{align}
with $k,m\in \F_{p^n}$, where $\ket{e(l)} \in \mathcal{B}_C$ are the elements of the computational basis (in arbitrary order).

Since for $n=1$ it holds that $\tr(k l^2+ml)=k l^2+ml$ the complete set of MUBs is given, apart form the computational basis, by the bases $\mathcal{B}_{k}$ containing the vectors
\begin{align}
\label{n1case}
\ket{v_k(m)}=\frac{1}{\sqrt{p}} \sum_{l=0}^{p-1} \omega_p^{kl^2+ml}\ket{e(l)} \ ,
\end{align}
where $k,m\in \mathbb{Z}_p$, and $\omega_p=e^{2\pi \mathbbm{i}/p}$ \cite{Iv}. The bases are clearly mutually unbiased, since
\begin{align}
\label{oddprime}
\left| \sum_{l=0}^{p-1}  \omega_p^{(k l^2 +m l)} \right| =\sqrt{p} \ ,
\end{align}
for any $k\neq 0 \ (\mathrm{mod} \ p)$ and $p$ an odd prime, which is known as a quadratic Gauss sum.

In \cite{wootters} the relation in Eq.~(\ref{charsum}) has been rewritten to avoid the use of the trace operator (which is generally hard to evaluate), and to generalize the construction to powers of two. As mentioned before, we can represent any element in $\F_{p^n}$ by an $n$--dimensional vector. More precisely, we choose a basis $\{b_i\}_{i=1}^{n}$, in $\F_{p^n}$ (for instance $\{1,\alpha,\alpha^2,\ldots,\alpha^{n-1} \}$, where $\alpha$ denotes a root of an irreducible polynomial) and write $l=\sum_i l_i b_i$. The vector associated with the polynomial $l$ is then $\vec{l}=(l_1,\ldots l_n)^T \in \mathbb{Z}_p^n$. The basic idea in rewriting Eq.~(\ref{charsum}) was to exploit the fact that the trace operator is linear. Therefore, $\tr(k l^2+ml)$ can be rewritten as $\vec{l}^T \left(\sum_{i=1}^{n} k_i M^{(i)} \right) \vec{l}+\vec{m}^T\vec{l},$ where $M^{(i)}$ are $n\times n$ symmetric matrices whose components $M_{u,v}^{(i)}$ are defined by the relation \bea \label{Eq_alpha} b_u b_v=\sum_{i=1}^n M_{u,v}^{(i)}b_i,\eea and the $n$--dimensional vector $\vec{m}$ is defined by $\tr(ml)=\sum_i m_i l_i$, and $k_i=\tr(k b_i)$. The complete set of MUBs is then given, apart form the computational basis, by ${\cal B}_{\vec{k}}=\{\ket{v_{\vec{k}}(\vec{m})} \}_{\vec{k}\in \mathbb{Z}_p^n}$, with
\bea \ket{v_{\vec{k}}(\vec{m})}=\frac{1}{\sqrt{d}} \sum_{\vec{l} \in \mathbb{Z}_p^n} \omega_p^{\vec{l}^T S_{\vec{k}} \vec{l}+\vec{m}^T\vec{l}} \ket{e(\vec{l})},\eea
where $S_{\vec{k}} \equiv \sum_{i=1}^{n} k_i M^{(i)}$. Note that the basis vectors $ \ket{v_{\vec{k}}(\vec{m})}$ are uniquely defined solely by the $n\times n$ matrices $M^{(i)}$. The requirement that the bases are mutually unbiased for all $p^n$ different values of $\vec{k}$, has then been shown without using the relation in Eq.~(\ref{charsum}). This is achieved by first noting that any symmetric matrix, as $M^{(i)}$, is diagonalizable in case $p$ is odd \cite{Ne72}), i.e. $M^{(i)}=P^{(i)} D^{(i)}  (P^{(i)})^T$, for any $i$, where $P^{(i)}$ is invertible and $D^{(i)}$ is diagonal and by proving that any non--trivial linear combination of these matrices, i.e. $S_{\vec{k}} \equiv \sum_{i=1}^{n} k_i M^{(i)}$ where $\vec{k}\neq 0$, is invertible over $\mathbb{Z}_p$. This implies that
the expression for the scalar product can then be written as an $n$--fold product of Eq.~(\ref{n1case}) \cite{wootters}. Note that from this result it follows that
for any non--singular symmetric matrix $S$ over $\mathbb{Z}_p$, i.e. $\det{S} \neq 0 \ \mathrm{mod} \ p$, it holds that
\begin{align}
\label{matrixtrace}
\left| \sum_{\vec{l} \in \mathbb{Z}_p^n} \omega_p^{\vec{l}^T S \vec{l} +\vec{m}^T \vec{l}} \right|=\sqrt{p^n} \ .
\end{align}

Note that for dimensions $d=2^n$, Eq.~(\ref{charsum}) does not hold. In fact, the absolute value would vanish, as can be easily verified. This prevents a straightforward generalization of the construction explained above to this case. In order to overcome this problem, the fourth root of unity, $\omega_4=\mathbbm{i}$, has been used in \cite{wootters}. In this way, a result similar to Eq.~(\ref{matrixtrace}) has been obtained. Namely,
\begin{align}
\left| \sum_{\vec{l} \in \mathbb{Z}_2^n} \mathbbm{i}^{\vec{l}^T S_{\vec{k}} \vec{l}} (-1)^{\vec{m}^T \vec{l}}  \right|=\sqrt{2^n} \ ,
\end{align}
where the sum runs over all elements $\vec{l}$ of $\mathbb{Z}_2^n$. Herein, $S_{\vec{k}}$ is again any non--trivial combination of the matrices $M^{(i)}$, as defined in Eq.~(\ref{Eq_alpha}). The crucial property of the matrices $S_{\vec{k}}= \sum_{i=1}^{n} k_i M^{(i)}$ is that they are symmetric $n \times n$ and have an odd determinant for all non--zero $\vec{k} \in \mathbb{Z}_2$. Using this result, a complete set of MUBs for $d=2^n$ has been constructed similar to Eq.~(\ref{oddcase}). Namely, the computational basis together with the bases $\mathcal{B}_{\vec{k}}$, where $\vec{k} \in \mathbb{Z}_2^n$, defined by the vectors
\begin{align}
\label{evencase}
\ket{v_{\vec{k}}(\vec{m})}=\frac{1}{\sqrt{d}} \sum_{\vec{l} \in \mathbb{Z}_2^n}  \mathbbm{i}^{\vec{l}^T S_{\vec{k}} \vec{l}} (-1)^{\vec{m}^T \vec{l}}   \ket{e(\vec{l})} \ ,
\end{align}
where $\vec{k}, \vec{m} \in \mathbb{Z}_2^n$, and each vector $\ket{e(\vec{l})}$ corresponds to an element of the computational basis $\mathcal{B}_C$.

Thus, summarizing this construction, one chooses a basis, $\{b_i\}_{i=1}^n$ of $\F_{p^n}$ and determines the symmetric matrices $M^{(i)}$ according to Eq.~(\ref{Eq_alpha}). The MUBs are then given, apart from the computational basis, as in Eq.~(\ref{oddcase}) for $p\geq 3$ and Eq.~(\ref{evencase}) for $p=2$ respectively.

\subsection{MUBs from maximally commuting bases} \label{constrbandyo}
Another construction of complete sets of MUBs for prime power dimensions was presented in Ref.~\cite{Bandyopadhyay}. Consider a complex Hilbert space of dimension $d$ (not necessarily a prime power), i.e. $\mathcal{H}=\mathbb{C}^d$. First note that the maximal number of pairwise orthogonal commuting unitary matrices $\{ U_i \}$ acting on $\mathbb{C}^d$ is $d$, which can be easily verified since the matrices are diagonal in the same basis. Let ${\cal M}=\{U_1,\ldots, U_{d^2}\}$ be an orthonormal basis of unitaries of the operator space, where w.l.o.g. $U_1=\one_d$. The set ${\cal M}$ is called a \emph{maximally commuting basis} if it can be partitioned as ${\cal M}=\{\one\}\bigcup{\cal C}_1 \ldots \bigcup{\cal C}_{d+1}$, where each class ${\cal C}_i$ contains $d-1$ commuting unitaries \footnote{Note that this is the maximum number since $\{\one\}\bigcup{\cal C}_i$ is a set of $d$ commuting orthogonal unitaries.}. In Ref.~\cite{Bandyopadhyay}, it was shown that if there exists such a maximal commuting basis of orthogonal unitary $d\times d$--matrices, then there exists a complete set of MUBs. The MUBs are simply the common eigenbases of the commuting operators within each class ${\cal C}_i$.

In order to construct complete sets of MUBs for prime power dimensions one can make use of the generalized Pauli operators \cite{Bandyopadhyay}. For a Hilbert space $\mathbb{C}^p$, these are defined as
\begin{align} \label{paulidefX}
X &= \sum_{k=0}^{p-1} \ket{(k+1) \ \mathrm{mod} \ p} \bra{k} \ , \\
Z &= \sum_{k=0}^{p-1} \omega_p^k \ket{k} \bra{k} \ , \label{paulidefZ}
\end{align}
where $\omega_p=e^{2\pi \mathbbm{i}/p}$. In the following we call a prime-dimensional quantum system a \emph{qupit}, in order to stress the difference to a qudit, which can have arbitrary dimension.  As can be easily seen, for prime dimension, i.e. $d=p$, the eigenbases of the $p+1$ operators, $Z, X , XZ, \ldots , X Z^{p-1}$, form a complete set of MUBs. For the more general case of prime powers, $d=p^n$, the generalized Pauli group on the Hilbert space, $\mathcal{H}=\mathbb{C}^d \simeq (\mathbb{C}^p)^{\otimes n}$, is generated by the set of operators
\begin{align}
P(\vec{k},\vec{l},\vec{m})=U(k_1,l_1,m_1) \otimes \cdots \otimes U(k_n,l_n,m_n) \ ,
\end{align}
wherein the operators $U(k_i,l_i,m_i)$, acting on system $i$ are of the form
\begin{align}
U(k,l,m)=\omega_p^k X^l Z^m \hspace{0.6cm} \mbox{where} \ \  k,l,m \in \mathbb{Z}_p \ ,
\end{align}
and the $n$--dimensional row vectors $\vec{k},\vec{l},\vec{m}$ are an abbreviation for the exponents, e.g. $\vec{k}=(k_1,\ldots,k_n)$. It can be straightforwardly shown that two elements of the Pauli group commute, i.e. $[P(\vec{k},\vec{l},\vec{m}), P(\vec{k'},\vec{l'},\vec{m'})]=0$, iff
\begin{align}
\label{commrelation}
\vec{l}\cdot \vec{m}^\prime- \vec{m}\cdot \vec{l}^\prime=0 \ (\mbox{mod} \ p) \ .
\end{align}
Moreover, two operators $P(\vec{k},\vec{l},\vec{m})$ and $P(\vec{k}',\vec{l}',\vec{m}')$ for which the corresponding $2n$--dimensional vectors $(\vec{l} , \vec{m})$ and $(\vec{l}' , \vec{m}')$ do not coincide are always mutually orthogonal. The class ${\cal C}_j$ is then defined via the $n$ operators $S^j_i=P(0,\vec{e}_i,\vec{m}_i^j)$ wherein $\vec{e}_i$ denotes the $i$'th unit vector, and the vectors $\vec{m}_i^j$ are to be determined. For conciseness, the $2n$--dimensional row vectors $(\vec{e}_i | \vec{m}_i^j)$ may be gathered in an $n\times 2n$ matrix for each $j$ . In this way, one obtains matrices of the form $E^j=(\one_n ,  A^j)$. Using the condition above, Eq.~(\ref{commrelation}), one finds that the $n$ Pauli operators $S^j_i$ commute for any fixed $j$ if the corresponding matrix $A^j$ is symmetric. The class ${\cal C}_j$ is then the set of all possible products of the generators, $S_i^j$ (excluding the identity), which are clearly all mutually commuting. Note that the multiplication of the operators amounts to the summation of the corresponding row vectors in $E^j$ modulo $p$. Thus, the common eigenbases of the operators in ${\cal C}_j$ are mutually unbiased if the corresponding generators are mutually orthogonal, which they are, as long as they are mutually independent. That is, the condition of mutual unbiasedness is that non of the generators $S_i^j$ can be written as a product of the operators from the other sets $\{S_i^k\}_{i=1}^n$ with $k\neq j$. This last condition is equivalent to the condition that there exists no non--zero $n$--dimensional (row) vector $\vec{v}$ such that $\vec{v} A_j=\vec{v} A_k$ for $k\neq j$. Hence, a sufficient condition for this independency is that $\det (A_j-A_k) \neq 0 \ (\mathrm{mod} \ p)$ for all $j\neq k$. Since this is exactly the same condition as the one required in the construction of MUBs presented in Ref.~\cite{wootters}, a possible choice is $A_k = \sum_{i=1}^{n} k_{i} M^{(i)}$, with $\vec{k}=(k_1,\ldots,k_n) \in \mathbb{Z}_p^n$ and $M^{(i)}$ defined in Eq.~(\ref{Eq_alpha}). In this way, it has been shown that there exists a maximally commuting basis ${\cal M}=\{\one\}\bigcup{\cal C}_1 \ldots \bigcup{\cal C}_{d+1}$ for all prime power dimensions. The complete set of MUBs are simply the common eigenbases of the generators $\{S_i^k\}_{i=1}^n$ of the stabilizer corresponding to each matrix $A_k$.

\section{Generalized graph--states} \label{SECGraph}
Graph--states, as the name indicates, are states which are characterized by mathematical graphs, i.e. a set of vertices and edges. The edges of a graph are gathered in a so-called adjacency matrix whose dimension corresponds to the number of vertices. There are two mathematically equivalent characterizations of graph--states \cite{Hein}. The first is the \emph{interaction picture}. It tells us how the generalized graph--states are constructed for a given adjacency matrix, by applying a particular class of $1$-- and $2$--body phase gates. The second is the \emph{stabilizer picture}. Here, a graph--state is uniquely defined via a set of operators --- the generators of a stabilizer --- which are elements of the Pauli group. In particular, the graph--state is defined as the unique eigenstate with eigenvalue one of all these operators.

\subsection{Definition}
\begin{figure}[h]
\includegraphics[height=3.3cm]{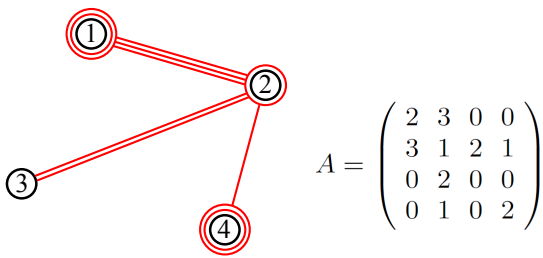}
\caption{(Color online) An example of a generalized multigraph and its associated adjacency matrix $A$. Edges between different vertices \textcircled{\footnotesize{i}} are represented by (red) lines, and self-loops by (red) circles. For instance, the vertex \textcircled{\footnotesize{1}} has three outgoing edges and two self-loops.}\label{multigraph}
\end{figure}

Let $G=(V,E)$ be an undirected graph with $n$ vertices $V=\{v_1,\ldots,v_n\}$ and a multiset $E=\{e_{i,j}\}$ of edges $e_{i,j}=(v_i,v_j)$. For our purpose, we permit multiple edges as well as self-loops, i.e. an edge $e_{i,j}$ may occur several times in $E$, and self-connections of the form $e_{i,i}$ are also allowed. Analogously to the case of simple graphs, such an undirected \emph{generalized multigraph} may be represented by a symmetric $n\times n$ matrix $A$, the adjacency matrix, where an entry $A_{i,j}=A_{j,i}$ corresponds to the number of edges $e_{i,j}$ between the vertices $v_i$ and $v_j$ (see also Refs.~\cite{labeledqubits,quditgraphs1,quditgraphs2}). In particular, the diagonal entries $A_{i,i}$ represent self-loops, and if two nodes $v_i$ and $v_j$ are not connected then $A_{i,j}=0$ (see Fig.~\ref{multigraph} as an example).

Consider an $n \times n$ adjacency matrix $A$ with entries in $\mathbb{Z}_p=\{0,\ldots,p-1\}$, where $p$ is a prime number. Given this matrix, a generalized graph--state is defined as follows. To each of the $n$ vertices, we associate a corresponding Hilbert space $\mathbb{C}^p$, with the standard basis $\{ \ket{0}, \ldots, \ket{p-1} \}$. Let the state $\ket{+} \in \mathbb{C}^p$ be the equally-weighted superposition of all basis states, i.e.
\begin{align}
\ket{+}= \frac{1}{\sqrt{p}} \sum_{i=0}^{p-1} \ket{i} \ .
\end{align}
Furthermore, we define the one-qupit phase operators
\begin{align}
\label{onequpitop}
U_{i,i} \, = \,\left\{\begin{array}{lll} \ \sum_{k=0}^{1} \omega_4^k \ket{k}\bra{k}_i  \hspace{1.5cm} & \mbox{for} \ p=2 \\
 \\
		\ \sum_{k=0}^{p-1} \omega_p^{k(k-1)/2} \ket{k}\bra{k}_i  & \mbox{for} \ p\geq3 \ , \end{array}\right.
\end{align}
and the two-qupit controlled-phase operator
\begin{align}
\label{twoqupitop}
\left. \begin{array}{lll} U_{i,j} & = \sum_{k,l=0}^{p-1} \omega_p^{kl} \ket{k}\bra{k}_i  \otimes \ket{l}\bra{l}_j \\
 &\hspace{5.0cm} \mbox{for} \ i \neq j \ , \\
& = \sum_{k=0}^{p-1}\ket{k}\bra{k}_i \otimes Z_j^k  \ .  \end{array} \right.
\end{align}
Here, and in the following, we use the notation $\omega_p=e^{2 \pi \mathbbm{i}/p}$, and $Z$ denotes the Pauli operator (local phase gate) as defined in Eq.~(\ref{paulidefZ}), where the index $i$ refers to the system the operator is acting on, e.g.
\begin{align}
\ket{k}\bra{k}_i=\mathbbm{1}\otimes \ldots \otimes \mathbbm{1} \otimes \underbrace{\ket{k}\bra{k}}_{i\mathrm{'th \ qupit}} \otimes \mathbbm{1} \otimes \ldots \otimes \mathbbm{1} \ .
\end{align}
Note that all phase operations $U_{i,j}$ and $Z_k$ commute $\forall i,j,k$ as they are diagonal in the computational basis.

For a given adjacency matrix $A$ with entries $A_{i,j}$, we define the generalized graph--state via the above operations as
\begin{align}
\label{graphstateconstr}
\ket{G}= \prod_{i \leq j} U_{i,j}^{A_{i,j}} \ket{+}^{\otimes n} \ .
\end{align}
Note that this is the standard description of graph--states \cite{Hein,quditgraphs1} which makes use of the $2$--body interactions given in Eq.~(\ref{twoqupitop}), extended by the local unitaries given in Eq.~(\ref{onequpitop}). The resulting states are also called \emph{labeled graph--states} \cite{labeledqubits,quditgraphs2}. Let us note that the operations $U_{i,i}$, which may be regarded as self-controlled phase gates, are local unitary operators which do not affect the entanglement properties of a graph--state.

For any Hilbert space $\mathcal{H}= \left(\mathbb{C}^p \right)^{\otimes n}$, one can construct an orthonormal basis in terms of graph--states. Namely, we define the \emph{graph--state basis} $\mathcal{B}_G=\{ \ket{G(m_1,\ldots,m_n)} \}_{m_i\in \mathbb{Z}_p}$ via
\begin{align}
\label{gengraphstate}
\ket{G(m_1,\ldots,m_n)}= Z^{m_1} \otimes \ldots \otimes Z^{m_n}\ket{G} \ ,
\end{align}
where $Z$ denotes the generalized Pauli operator as defined in Eq.~(\ref{paulidefZ}). All basis states $\ket{G(m_1,\ldots,m_n)}$ are local-unitarily equivalent since each $Z$ acts locally.

Consequently, each graph (i.e. adjacency matrix) corresponds to a basis of the Hilbert space. The following construction of MUBs is based on these particular bases. That is, each basis of a set of MUBs will be represented by a single graph. In this context, it should be noted that the diagonal entries of the adjacency matrix for qubits ($p=2$) can be treated modulo $2$, even though the local phase in $U_{i,i}$ is $\omega_4$. This is because a change of the entry $A_{i,i}$ from $2$ to $0$ (or, $3$ to $1$), results in the same basis $\mathcal{B}_G$ but with permuted basis elements [$m'_i=(m_i+1) \ \mathrm{mod} \ 2$], as for $p=2$ it holds that $U_{i,i}^2=Z_i$.

\subsection{Stabilizers of generalized graph--states}
As mentioned above, generalized graph--states can be characterized in terms of stabilizers from the Pauli group, which can be determined straightforwardly. In particular, a graph--state $\ket{G(m_1,\ldots,m_n)}$, corresponding to the adjacency matrix $A$, is stabilized by a group of operators which is defined by $n$ generators, $\{S_i\}_{i=1}^n$. The graph--state $\ket{G}=\ket{G(0,\ldots,0)}$ is the unique eigenstate of all $S_i$ to eigenvalue one. In Ref.~\cite{labeledqubits}, it was shown that, for $p=2$, any graph--state $\ket{G(m_1,\ldots,m_n)}$ defined by the adjacency matrix $A$ satisfies
  \begin{align}
S_i \ket{G(m_1,\ldots,m_n)} = \omega_2^{m_i} \ket{G(m_1,\ldots,m_n)} \ ,
\end{align}
where
  \begin{align}
S_i= \left( \omega_4^{A_{i,i}} X_i Z_i^{A_{i,i}} \right) \bigotimes_{j \neq i} {Z_j}^{A_{i,j}} \ , \hspace{0.7cm} 1 \leq i \leq n \ .
\end{align}
Similarly, for $p\geq3$, it was shown (see Ref.~\cite{quditgraphs2}) that any graph--state $\ket{G(m_1,\ldots,m_n)}$ defined by the adjacency matrix $A$ satisfies
      \begin{align}
S_i \ket{G(m_1,\ldots,m_n)} = \omega_p^{-m_i} \ket{G(m_1,\ldots,m_n)} \ ,
\end{align}
where
      \begin{align}
      S_i= \left( X_i Z_i^{A_{i,i}} \right) \bigotimes_{j \neq i} {Z_j}^{A_{i,j}} \ , \hspace{0.5cm} 1 \leq i \leq n \ .
\end{align}
\section{Mutual unbiasedness of graph--states} \label{SECMUBGraph}
In the following sections we present a novel formalism that allows us to attain mutual unbiasedness (MU) between pairs of graph--state bases. Instead of starting with condition Eq.~(\ref{charsum}), we consider the overlap of pairs of generalized graph--states. Using some of the concepts given in Ref.~\cite{wootters}, we rederive a sufficient condition for mutual unbiasedness from Sec.~\ref{woottconstr} and Sec.~\ref{constrbandyo}, which allows us to establish its connection to the adjacency matrices of generalized graph--states. In this way, we obtain a simple and insightful graphical representation of MUBs. We start out by deriving the condition for mutual unbiasedness in the case of a single qupit (Sec.~\ref{MUsingle}) and two qupits (Sec.~\ref{MUtwo}). Here, we will only need the well-known \emph{orthogonality relation}
\begin{align}
\label{ortho}
\sum_{l=0}^{p-1} \omega_p^{kl} = \delta_{k,0} p \ ,
\end{align}
in order to prove that certain states are mutually unbiased. Those results will then be combined in Sec.~\ref{MUmany} to derive the conditions for MU for multipartite states. Note that the following arithmetics in the exponent of $\omega_p$ are to be read modulo $p$, since it holds that $\omega_p^{k}=\omega_p^{k+p}$ for any exponent $k$ of $\omega_p$.

\subsection{Mutual unbiasedness for a single qupit}
\label{MUsingle}
Consider the $1 \times 1$ adjacency matrix $A=(A_{1,1})$ over $\mathbb{Z}_p$. In the following we use the abbreviation $r\equiv A_{1,1}$. Let us begin by showing that the $p$ different one-qupit graph--states $\ket{G_r}= U_{1,1}^{r} \ket{+}$ with different $r \in \mathbbm{Z}_p$ and associated bases $\mathcal{B}_r=\{ \ket{G_r(m_1)} \}_{m_1\in \mathbbm{Z}_p}$ are mutually unbiased, i.e. for any pair $r, r' \in \mathbbm{Z}_p$ with $r\neq r'$ it holds that
\bea \label{EqH1} H_1=|\bra{G_{r'}(m'_1)} G_{r}(m_1) \rangle|^2=\frac{1}{p} \ , \eea
for all $m_1,m'_1 \in \mathbbm{Z}_p$.

First, consider a single qubit and the quantity $H_1=|\bra{+} U_{1,1} Z_1^{m_1} \ket{+} |^2$, which corresponds to the overlap of an arbitrary pair of graph--states that differ by the local operations $U_{1,1}$ and an arbitrary $Z_1^{m_1}$ with $m_1 \in \mathbb{Z}_2$. It is straightforward to verify that $H_1=\frac{1}{2}$ holds for any $m_1 \in \mathbb{Z}_2$, since $U_{1,1}\ket{+}=\frac{1}{\sqrt{2}}(\ket{0}+\mathbbm{i} \ket{1})$ and $Z_1U_{1,1}\ket{+}=\frac{1}{\sqrt{2}}(\ket{0}-\mathbbm{i} \ket{1})$. Hence, this is simply a compact reformulation of the well-known fact that the bases
\begin{align}
\label{Xbasis}
\mathcal{B}_{0}&=\{ \frac{1}{\sqrt{2}} (\ket{0} + \ket{1}), \frac{1}{\sqrt{2}} (\ket{0} - \ket{1})\} \ ,\\
\mathcal{B}_{1}&=\{ \frac{1}{\sqrt{2}} (\ket{0} + \mathbbm{i} \ket{1}), \frac{1}{\sqrt{2}} (  \ket{0} - \mathbbm{i} \ket{1})\} \ ,
\label{Ybasis}
\end{align}
which are the normalized eigenvectors of the Pauli matrices $X$ and $Y$, are mutually unbiased.

Next, for a single qupit with $p \geq 3$, $H_1$ as given in Eq.~(\ref{EqH1}) can be written as $|\bra{+} U_{1,1}^{r-r'} Z_1^{m_1-m_1'} \ket{+} |^2$, where $D_{1,1} \equiv r-r' \neq 0$. Thus, in order to show that all $p$ different bases, $\mathcal{B}_{r}=\{ \ket{G_r(m_1)} \}_{m_1\in \mathbbm{Z}_p}$, for $r=0,\ldots , p-1$, are mutually unbiased, we show that $H_1=|\bra{+} U_{1,1}^{D_{1,1}} Z_1^{m_1} \ket{+} |^2= \frac{1}{p}$ for all $D_{1,1} \neq 0$. This can easily be shown as follows. Consider
\begin{align}
H_1&=\frac{1}{p^2} \left|\sum_{k=0}^{p-1} \omega_p^{ D_{1,1} 2^{-1} k(k-1) +m_1 k}\right|^2  \ ,
\end{align}
wherein $2^{-1}=\frac{p+1}{2} \in \mathbb{Z}_p$ denotes the multiplicative inverse of the element $2\in \mathbb{Z}_p$. Using the abbreviation $m'_1=m_1-2^{-1}D_{1,1}$, we have
\begin{align}
H_1&=\frac{1}{p^2}\left(\sum_{k=0}^{p-1}\omega_p^{2^{-1} D_{1,1}k^2+m'_1 k}\right) \left(\sum_{l=0}^{p-1}\omega_p^{-2^{-1}D_{1,1}l^2-m'_1 l}\right)  \nonumber \\
&=\frac{1}{p^2}\sum_{k,l=0}^{p-1}\omega_p^{(2^{-1}D_{1,1}(k+l)+m'_1)(k-l)} \ .
\end{align}
The last equation can be rewritten as
\begin{align}
H_1=\frac{1}{p^2} &  \left(  \left[\sum_{k=l}\omega_p^{(2^{-1}D_{1,1}(k+l)+m'_1)(k-l)}\right] \right. \nonumber  \\
   & + \left. \left[\sum_{k\neq l} \omega_p^{(2^{-1}D_{1,1}(k+l)+m'_1)(k-l)}\right] \right)  \ .
\end{align}
Here, the first of the two terms in square brackets is equal to $p$ since $k-l=0$. Substituting $(k-l)$ by $s \in \{ 1, \ldots , p-1 \}$, the second term can be written as $\sum_{s=1}^{p-1} \left( \sum_{l=0}^{p-1} \omega_p^{(2^{-1}D_{1,1}(2l+s)+m'_1)s} \right)$. As $p \geq 3$ is prime, the function $t:\mathbb{Z}_p \rightarrow \mathbb{Z}_p$,  $t(l)=2^{-1}D_{1,1}(2l+s)+m'_1$ is bijective for any $D_{1,1}\neq 0$. Hence, for any $s\neq 0$ (and any $D_{1,1}\neq 0$) we have $\sum_{l=0}^{p-1} \omega_p^{(2^{-1}D_{1,1}(2l+s)+m'_1)s}=\sum_{t=0}^{p-1} \omega_p^{t s}=0$; which implies that $H_1=\frac{1}{p}$, as claimed. Therefore, all bases $\mathcal{B}_{r}=\{ \ket{G_r(m_1)} \}_{m_1\in \mathbbm{Z}_p}$ are mutually unbiased for different values of $r \in \mathbb{Z}_p$. Here, recall that $r=A_{1,1}$. Note that the one-qupit phase operator $U_{i,i}$ was simply defined in accordance with the quadratic Gauss sum from Eq.~(\ref{oddprime}).

\subsection{Mutual unbiasedness for two qupits} \label{MUtwo}
We derive an analogous result for two qupits via the two-body phase gate defined in Eq.~(\ref{twoqupitop}). Namely, the $p$ different two-vertex graph--states $\ket{G_r}= U_{1,2}^{r} \ket{+}^{\otimes 2}$ with different $r \in \mathbbm{Z}_p$ and associated bases $\mathcal{B}_r=\{ \ket{G_r(m_1,m_2)} \}_{m_1,m_2\in \mathbbm{Z}_p}$ are mutually unbiased. More specifically, each $r$ corresponds to a graph--state basis defined by a $2 \times 2$ adjacency matrix with entries $A_{1,1}=A_{2,2}=0$ and $A_{1,2}=A_{2,1}=r$.

We first consider the quantity $H_{2}=|\bra{+}^{\otimes 2} U_{1,2} Z_1^{m_1} Z_2^{m_2} \ket{+}^{\otimes 2} |^2$, which corresponds to the overlap of a pair of two-qupit graph--states that differ by the entangling operation $U_{1,2}$, where $m_1, m_2 \in \mathbb{Z}_p$ is arbitrary. Explicitly, $H_2$ reads
\begin{align}
H_{2}=\frac{1}{p^4} \left|\sum_{k,l=0}^{p-1} \omega_p^{kl+m_1 k + m_2 l}\right|^2 \ .
\end{align}
Splitting the inner sum into two parts with $k=-m_2$ and $k\neq - m_2$, we obtain
\begin{align*}
\frac{1}{p^4} \left| \left[\sum_{l=0}^{p-1} \omega_p^{-m_1 m_2}\right] + \left[ \sum_{k\neq-m_2} \omega_p^{m_1 k} \left(\sum_{l=0}^{p-1} \omega_p^{(k+m_2)l}\right)\right] \right|^2 \ .
\end{align*}
Herein, the first of the two terms in square brackets is equal to $ p\times \omega_p^{-m_1 m_2}$, whereas the second term vanishes since for any $k$ satisfying $k+m_2 \neq 0$ it holds that $\sum_{l=0}^{p-1} \omega_p^{(k+m_2)l}=0$ [see Eq.~(\ref{ortho})]. Hence, in total we have $H_2=\frac{1}{p^4} \left| p \ \omega_p^{-m_1 m_2} \right|^2= \frac{1}{p^2}$. For $p$ prime, the same result is obtained for all non--zero powers $A_{1,2} \in \{ 1,\ldots, p-1\}$ of $U_{1,2}$ in $H_2$, as replacing the running index $k$ by any $k'=A_{1,2}k$ (in the sum which vanishes) clearly does not affect the result.

Thus, we have shown that for any pair of adjacency matrices of the form
\begin{align}
A=\left(\begin{array}{cc}
                                                                0 & r \\
                                                                r & 0 \\
                                                              \end{array}
                                                            \right) \ , \hspace{0.5cm}
A'=\left(
                                                              \begin{array}{cc}
                                                                0 & r' \\
                                                                r' & 0 \\
                                                              \end{array}
                                                            \right) \ ,
\end{align}
with $r \neq r' $, the corresponding graph--state bases $\mathcal{B}_{r}=\{ \ket{G_r(m_1,m_2)} \} $ and $\mathcal{B}_{r'}=\{ \ket{G_{r'}(m'_1,m'_2)} \}$ are mutually unbiased as $|\braket{G_{r'}(m'_1,m'_2)}{G_{r}({m_1,m_2})}|^2=|\bra{+}^{\otimes 2} C_{1,2}^{r-r'} Z_1^{m_1-m_1'} Z_2^{m_2-m_2'} \ket{+}^{\otimes 2} |^2=H_2=\frac{1}{p^2}$ for all $D_{1,2}=r - r'\neq 0$ and all $m_i,m_i^\prime \in \mathbb{Z}_p$ with $i=1,2$. %\hfill $\square$

\subsection{Mutual unbiasedness for several qupits} \label{MUmany}
We now combine the observations we have made for a single and a pair of qupits to construct MUBs for arbitrary multi-qupit systems. First, consider the general overlap \bea H_n&=&|\braket{G'({m'_1,\ldots,m'_n})}{G(m_1,\ldots,m_n)}|^2\\ \nonumber &=&|\bra{+}^{\otimes n} \prod_{i \leq j} U_{i,j}^{A_{i,j}-A'_{i,j}} \prod_{k=1}^{n} Z_k^{m_k-m_k'} \ket{+}^{\otimes n} |^2\eea of a pair of graph--states in $\mathcal{H}= \left(\mathbb{C}^p\right)^{\otimes n}$. First, note that the overlap $H_n$ factors into a product whenever the difference between the adjacency matrices, $D=A-A'$, is block diagonal. Second, according to the previous section, a $1 \times 1$ block $(D_{i,i}) \neq 0$ yields a factor $H_1=\frac{1}{p}$, and a $2\times 2$ block $\left(\begin{array}{cc} 0 & D_{j,k} \\ D_{j,k} & 0 \\ \end{array}\right)$ with $D_{j,k} \neq 0$ gives a factor $H_2=\frac{1}{p^2}$. Consequently, if the difference between the adjacency matrices, $D=A-A'$, is a direct sum of $1\times 1$ and $2 \times  2$ blocks of this kind, the overlap becomes $H_n= H_1^{h_1} H_2^{h_2}$, where $h_1$ and $h_2$ are the multiplicities of the corresponding blocks, where $h_1+2 h_2=n$. Hence, in total we get $H_n=\frac{1}{p^n}$ in this case, which means that the two bases are mutually unbiased.

We are going to show now that the sufficient condition that the difference between the adjacency matrices, $D=A-A'$ is block--diagonal as mentioned above is not necessary for the two states (and the corresponding bases) to be mutually unbiased. In fact, we will show that whenever $D$ is a symmetric $n \times n$ matrices with full rank (or equivalently, non--zero determinant in $\mathbb{Z}_p$), the corresponding graph--state bases are MUBs. In order to do so we treat the two cases, $p\geq 3$ [case (i)]  and $p=2$ [case (ii)] separately.

\emph{Case (i) ---} First, consider an arbitrary multi-qupit system $d=p^n$ with $p \geq 3$. Suppose $D$ has the required block structure, i.e. is a direct sum of $1\times 1$ and $2 \times 2$ regular blocks, such that
\begin{align}
H_n=\frac{1}{p^{2n}} \left|\sum_{k_1,\ldots,k_n=0}^{p-1}  \omega_p^{\sum_{l} 2^{-1} D_{l,l} k_l^2 + \sum_{i < j } D_{i,j} k_i k_j + \sum_{x} m_x k_x}\right|^2 \ ,
\end{align}
satisfies $H_n = \frac{1}{p^n}$, where the $m_x\in \mathbb{Z}_p $ are arbitrary. Using $\omega_p^{D_{i,j}k_i k_j}=\omega_p^{2 \times 2^{-1} D_{i,j}k_i k_j}$, we can write the sum in the exponent as a quadratic form, i.e.
\begin{align}
\label{quform}
H_n=\frac{1}{p^{2n}} \left|  \sum_{\vec{k}}  \omega_p^{2^{-1} \vec{k}^T D \vec{k} + \vec{m}^T \vec{k}}\right|^2 \ ,
\end{align}
where $\vec{k}=(k_1,\ldots,k_n)^T$.

Obviously, the overlap is invariant under reordering of the summation over $\vec{k}$. Changing the order of the summation is equivalent to a transformation $\vec{k} \rightarrow P \vec{k}$ using an invertible $n\times n$ matrix $P$ with entries in $\mathbb{Z}_p$. Inserting this in Eq.~(\ref{quform}) leads to a congruence transformation $P^T D P = D'$ (and $\vec{m}^{\prime}=P^T\vec{m}$). Therefore, one realizes that not only all matrices $D$ possessing the proper block structure lead to MU, but also all matrices $D'$ which are congruent to them. Note that these are simply all symmetric invertible matrices, since it has been shown that any symmetric matrices over $\mathbb{Z}_p$, with $p\geq 3$ can be transformed into a diagonal matrix via a congruence transformation \cite{AAlbert}.

\emph{Case (ii) ---} The same procedure can also be adapted to multi-qubits, i.e. to the case where $d=2^n$. There,
\begin{align}
\label{thesum}
H_n&=\frac{1}{2^{2n}} \left|\sum_{k_1,\ldots,k_n=0}^{1} \omega_4^{\sum_{l} D_{l,l} k_l} \omega_2^{\sum_{i < j} D_{i,j} k_i k_j + \sum_{x} m_x k_x}\right|^2 \ .
\end{align}
For $k_i$ in $\mathbb{Z}_2$ it holds that $k_i=k_i^2$ and $\omega_2=\omega_4^2$, and therefore we can write
\begin{align}
H_n&=\frac{1}{2^{2n}} \left|\sum_{k_1,\ldots,k_n=0}^{1} \omega_4^{\sum_{l} D_{l,l} k_l^2 + \sum_{i < j} 2 D_{i,j} k_i k_j + \sum_{x} 2 m_x k_x}\right|^2   \nonumber \\
&=\frac{1}{2^{2n}} \left|  \sum_{\vec{k}} \omega_4^{\vec{k}^T D \vec{k} +  2 \vec{m}^T \vec{k}}\right|^2 .
\label{EqQ},
\end{align}
Note that the matrix $D$ now can have entries in $\mathbb{Z}_4$ as the base of the exponent is $\omega_4$, which means that the arithmetics are to be done modulo $4$. However, in Eq.~(\ref{thesum}) we see that the off-diagonal elements $D_{i,j}$ can be treated modulo $2$, as they are actually exponents of $\omega_2=-1$. Furthermore, writing the diagonal elements $D_{l,l}$ as $D_{l,l}=o_l + 2 e_l$, where $o_l=D_{l,l}\mod 2$ with $o_l, e_l \in \mathbb{Z}_2$, we see that the even parts, $e_l$, of the diagonal elements can always be shifted into the vector $\vec{m}$. That is, we can rewrite Eq.~(\ref{EqQ}) as
\begin{align*}
H_n&=\frac{1}{2^{2n}} \left|\sum_{k_1,\ldots,k_n=0}^{1} \omega_4^{\sum_{l} o_{l,l} k_l} \omega_2^{\sum_{j > i} D_{i,j} k_i k_j + \sum_{x} (m_x+e_x) k_x}\right|^2 \ ,
\end{align*}
which means that the vector $\vec{m}$ changes to $\vec{m}'=\vec{m}+\vec{e}$. As the odd parts $o_l$ are simply $o_l=(D_{l,l} \ \mathrm{mod} \ 2)$, we finally conclude that all entries of $D$ may be treated modulo $2$. Therefore, similarly to the qupit case, where $p\geq3$, one realizes that all matrices $D'$ over $\mathbb{Z}_2$ which are congruent to $D=(1)^{\oplus h_1}  \oplus \left( \begin{array}{cc} 0 & 1 \\ 1 & 0 \\ \end{array} \right)^{\oplus h_2}$ with $h_1+2 h_2=n$ give rise to MU. Again, a matrix fulfills this condition iff it is a symmetric $n\times n$ matrix with full rank, i.e. the determinant is one ($\mathbb{Z}_2$) \cite{AAlbert}.

Thus, we have shown that for any $p$ and $n$, the overlap as given in Eq.~(\ref{quform})   [case(i)] or in Eq.~(\ref{EqQ}) [case(ii)] equals $1/p^{n}$ if $D$ is congruent (in $\mathbb{Z}_p$) to $D  \oplus \left( \begin{array}{cc} 0 & 1 \\ 1 & 0 \\ \end{array} \right)^{\oplus h_2}$, where $D$ is a $h_1\times h_1$ diagonal matrix with non--zero diagonal elements in $\mathbb{Z}_p$ and $h_1+2 h_2=n$. Due to an established result of matrix analysis \cite{AAlbert}, those matrices are easily characterized, since $D$ over $\mathbb{Z}_p$ fulfills the above condition iff $D$ is a symmetric $n \times n$ matrix with full rank --- or equivalently, non--zero determinant over $\mathbb{Z}_p$. Let us summarize this fact in the following lemma (see also Refs.~\cite{wootters,Bandyopadhyay} and Sec.~\ref{SECConstr}).

\begin{lemma} \label{MUBcondition} Let $A_r$ and $A_s$ be a pair of symmetric $n \times n$ matrices over $\mathbbm{Z}_p$. If it holds that
\bea \det (A_r -A_s) \neq 0 \ \mathrm{mod} \ p  \ , \eea
then the graph--state bases [see Eq.~(\ref{gengraphstate})] corresponding to the adjacency matrices $A_r$ and $A_s$ are mutually unbiased. \end{lemma}

\section{Complete sets of mutually unbiased bases} \label{SECCompMUBs}
\begin{figure}[htb]
\includegraphics[width=5.5cm]{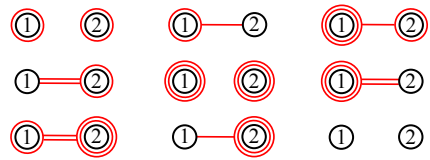}
\caption{(Color online) A complete set of graph--state MUBs for two qutrits generated by the vector $\vec{d}=(1,0)$ defining a symmetric tridiagonal matrix $Q$, as given in Eq.~(\ref{tridiag}), whose characteristic polynomial $f(x)=\mathrm{char} (Q)$ is irreducible. Note that the first two graphs in the picture are fundamental, i.e. all others are linear combinations of them over $\mathbb{Z}_3$.}\label{2qutritsgraph} %vector (10) poly x^2+2x+2
\end{figure}

 Now, we exploit these results to construct complete sets of $d+1$ MUBs for arbitrary dimensions, $d=p^n$. First, notice that any graph--state as defined in Eq.~(\ref{graphstateconstr}) and Eq.~(\ref{gengraphstate}) is always mutually unbiased with respect to the computational basis. Therefore, a set of $p^n$ mutually unbiased graph--state bases is tantamount to a complete set of $p^n+1$ MUBs \footnote{For instance, for one qupit, the $p$ single vertex graph--state bases $\{ \ket{G_r(m_1)} \}_{m_1\in \mathbbm{Z}_p}$ directly give rise to a complete set.}. Moreover, each basis is obtained from a single graph--state by applying local $Z$ operations [see Eq.~(\ref{gengraphstate})]. For instance, the $9$ multigraphs in Fig.~\ref{2qutritsgraph} correspond to a complete set of $9+1=10$ MUBs for the Hilbert space $\mathcal{H}=\mathbb{C}^9$. Note that the computational basis is never illustrated.

According to Lemma \ref{MUBcondition}, we need to find a set of $p^n$ adjacency matrices, $S=\{A_0, \ldots, A_{p^n-1} \}$, such that $\det ( A_{r}-A_{s}) \neq 0 \ \mathrm{mod} \ p$ for all $r\neq s$ [as was also required in the other approaches (see Sec.~\ref{SECConstr})]. If this condition is satisfied, then the graph--state bases corresponding to the adjacency matrices $\{A_0,\ldots,A_{p^n-1}\}$ form a complete set of MUBs. The existence of such matrices for all prime powers is already guaranteed by the results presented in \cite{wootters}. As mentioned in Sec.~\ref{IntroFF} it has been shown there that one possible choice would be the matrices $A_k = \sum_{i=1}^{n} k_{i} M^{(i)}$, where each $k$ corresponds to one of the $p^n$ possible settings of the vector $\vec{k}_i=(k_1,\ldots,k_n) \in \mathbb{Z}_p^n$, with the $n$ different symmetric $n\times n$ matrices $M^{(i)}$ as defined in Eq.~(\ref{Eq_alpha}).

Here, we present an alternative, constructive method which yields sets of matrices that satisfy the required condition. In contrast to the set of matrices $\{M^{(i)}\}_{i=1}^n$ we give a simple method to construct a single symmetric matrix, whose powers (and sums of powers) will lead to the desired set. Moreover, we will show that the complete set of MUBs can be encoded by a single $n$--dimensional vector.

To this end, we adopt concepts from the theory of finite fields and their representations \cite{Lidl}. For our construction we are going to exploit the simple observation that the difference $\delta=\alpha-\beta$ of any two unequal elements $\alpha,\beta \in \F_{p^n}$ of a finite field has a multiplicative inverse $\delta^{-1}$, since $\delta$ is a member of the multiplicative group $(\F_{p^n} \backslash \{0\}, \cdot)$. Suppose now that the set of symmetric $n\times n$ matrices $S=\{A_0, \ldots, A_{p^n-1} \}$ over $\mathbb{Z}_p$ was a matrix representation of $\mathbb{F}_{p^n}$ with respect to the ordinary matrix addition and matrix multiplication. In this case, all matrices $D_{r,s} = A_{r}-A_{s}$ would be invertible for $A_r \neq A_s$. Thus, the set $S$ would have the desired property.

We now discuss how such a representation may be obtained. Note that the following ideas are based on the matrix representation given in Ref.~\cite{Lidl}. Here, and in the following, we denote the $n\times n$ zero matrix by $\mathds{O}_n$, and the $n\times n$ identity matrix by $\mathbbm{1}_n$. Consider an $n \times n$ matrix $Q$ and the polynomials $\sum_{i} c'_i Q^i$, both over $\mathbb{Z}_p$. Let $f_m(x)=x^m+a_{m-1} x^{m-1}+\ldots + a_{0}x^0$ be the (monic) polynomial (over $\mathbb{Z}_p$) of minimal degree $m$ such that $f_m(Q)=\mathds{O}_n$. Then, as it holds that $Q^m=-a_{m-1} Q^{m-1}-\ldots - a_{0}Q^0$, any polynomial $\sum_{i} c'_i Q^i$ of arbitrary degree equals a polynomial $\sum_{i=0}^{m-1} c_i Q^i$ of degree smaller than $m$. Therefore, there are only $p^m$ polynomials; namely the elements of the residue class $\F_p [Q] \backslash (f_m(Q))$ \footnote{These $p^m$ polynomials are indeed all different, as otherwise for a pair of polynomials with different cofficients, say $\sum_{i=0}^{m-1} {c_i} Q^i$ and $\sum_{i=0}^{m-1} {c_i'} Q^i$, one could achieve that $\sum_{i=0}^{m-1} (c_i-c_i') Q^i=\mathds{O}_n$, which would be in contradiction with $f_m(x)$ being the polynomial of minimal degree with this property.}. As $\F_p [Q] \backslash (f_m(Q))$ is isomorphic to $\F_p [x] \backslash (f_m(x))$, we have that if $f_m(x)$ is of degree $m=n$ and irreducible over $\mathbb{Z}_p$, then $\F_p [Q] \backslash (f_m(Q))$ represents the finite field $\F_{p^n}$, as discussed in Sec.~\ref{IntroFF}. In order to achieve this, it suffices to choose $Q$ such that its characteristic polynomial $f_c(x)=\mathrm{char}(Q)=\det{(x \mathbbm{1} - Q)}$ is irreducible, as in this case it automatically holds that $f_m(x)=f_c(x)$ with polynomial degree $\deg(f_m(x))=n$ \footnote{Due to the Cayley-–Hamilton theorem it holds that $f_c(Q)=\mathds{O}_n$, where $f_c(x)$ denotes the characteristic polynomial of the $n \times n$ matrix $Q$, i.e. $f_c(x)=\mathrm{char}(Q)=\det{(x \mathbbm{1} - Q)}$. Furthermore, by definition it also holds that $f_m(Q)=\mathds{O}_n$ where $\deg (f_m(x))= m \leq n$. One can show that $f_m(x)$ is always a factor of $f_c(x)$, i.e. $f_c(x)=q(x) f_m(x)$ [see e.g. Ref.~\cite{Perlis}]. However, if the characteristic polynomial $f_c(x)$ is irreducible, then there is only the trivial factorization $f_c(x)=q(x) \cdot f_m(x)$ with $q(x)=1$; and hence, the polynomial of minimal degree $m$ such that $f_m(Q)=\mathds{O}_n$ is the characteristic polynomial itself.}. Therefore, if the characteristic polynomial of $Q$ is irreducible over $\mathbb{Z}_p$, then the set $\{Q^i\}_{i=0}^{n-1}$ forms a basis of the representation of $\F_{p^n}$. We state this fact in the following lemma.

\begin{lemma} Let $Q$ be an $n \times n$ matrix over $\mathbbm{Z}_p$ whose characteristic polynomial is irreducible. Then, the polynomials in $Q$ over $\mathbbm{Z}_{p}$ of degree less than $n$, i.e.
\bea S=\{\sum_{i=0}^{n-1} a_i Q^i, \vec{a}=(a_0,\ldots,a_{n-1})\in \mathbbm{Z}_p^n\} \ ,\eea
are a matrix representation of $\F_{p^n}$, with respect to matrix addition and matrix multiplication. \end{lemma}

 A further property that we can exploit is that the multiplicative group $(\F_{p^n}\backslash \{0\}, \cdot)$ is cyclic \cite{Lidl}. Hence, there always exists a primitive element, which generates the whole group (apart form the $0$ element). In case the matrix $Q$ constitutes a primitive element, then any non--zero element of $\{\sum_{i=0}^{n-1} a_i Q^i, \vec{a}=(a_0,\ldots,a_{n-1})\in \mathbbm{Z}_p^n\}$ is a power $Q^i$. This leads to the following

\begin{corollary} Let $Q$ be an $n \times n$ matrix over $\mathbbm{Z}_p$ whose characteristic polynomial is a primitive polynomial. Then, the powers of $Q$ of degree less than $p^n-2$, i.e.
\bea \{Q^i\}_{i=0}^{p^n -2},\eea
are a representation of $\F_{p^n}\backslash \{0\}$.
\end{corollary}

In order to obtain the representation of $\F_{p^n}$ one simply has to include the $n\times n$ zero--matrix, $\mathds{O}_n$, i.e. $S=\{Q^i\}_{i=0}^{p^n -2} \bigcup \{ \mathds{O}_n \}$. Since this is a matrix representation of $\F_{p^n}$, all matrices corresponding to a difference of those matrices are invertible. However, in order to find the desired set $S$, it remains to show that we can always find a \emph{symmetric} matrix $Q$ whose characteristic polynomial is irreducible.

In the subsequent sections we show that a matrix representation of $\F_{p^n}$ in terms of symmetric $n \times n$ matrices $S=\{A_0, \ldots, A_{p^n-1} \}$ over $\mathbb{Z}_p$ indeed always exists. Moreover, we will present two constructive methods of finding the single matrix $Q$ required to construct the set $S$. Whereas the first method is proven to work in general, i.e. for $p$ and $n$ arbitrary, the second is proven to work only for multipartite qubits, i.e. $p=2$. However, numerically we observe that this method also works for other values of $p$. The advantage of the second method is that a complete set of MUBs can be presented in a single $n$--dimensional vector.
\begin{figure}[h!]
\includegraphics[width=8cm]{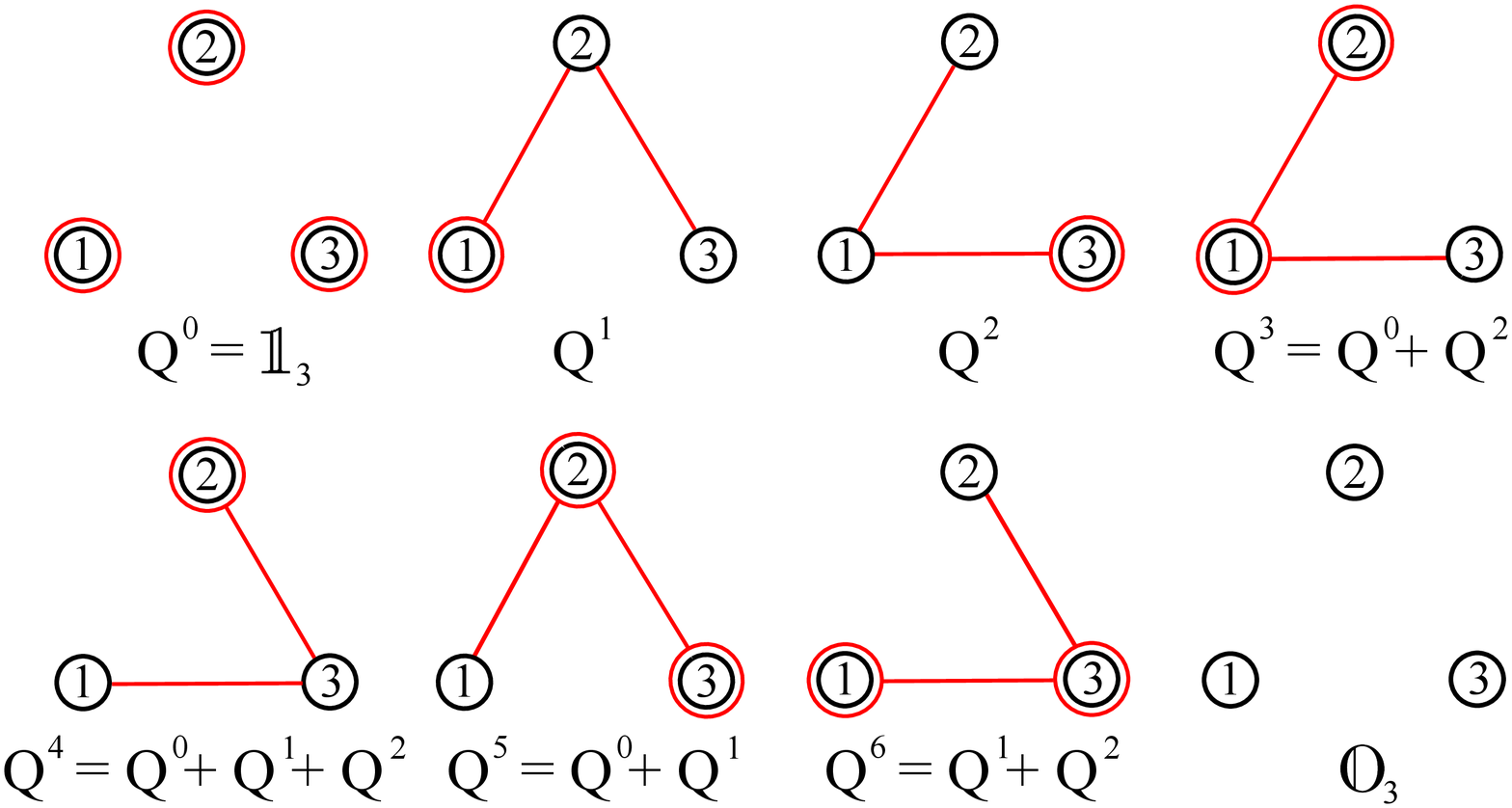}
\caption{(Color online) A complete set of graph--state MUBs for three qubits resulting from the vector $\vec{d}=(1,0,0)$ defining a tridiagonal matrix $Q$, as given in Eq.~(\ref{tridiag}), whose characteristic polynomial $f(x)=\mathrm{char} (Q)=x^3+x^2+1$ is irreducible. Below each graph we give its adjacency matrix. Note that any of the eight adjacency matrices is a linear combination (over $\mathbb{Z}_2$) of the first three adjacency matrices, which are the powers $0$, $1$ and $2$ of the matrix $Q$. In a graphical sense, this means that by overlaying any two graphs in the picture, we get another graph from the set. Here, overlaying (i.e. superimposing) amounts to summing up the (red) lines, modulo $p$ (which is $2$ in this case). Note further that the set of possible linear combinations also includes the zero matrix $\mathds{O}_3$, which corresponds to a graph--state basis Eq.~(\ref{gengraphstate}) defined by $\ket{G}=\ket{+}^{\otimes 3}$. As $f(x)$ is also a primitive polynomial, any non--zero adjacency matrix is a power of $Q$. The illustrated set has the following entanglement properties. As in the first and last graph all vertices are disconnected, the corresponding bases are fully separable. The six other graphs represent bases whose elements are local-unitarily equivalent to the $GHZ$--state $\ket{GHZ}=\frac{1}{\sqrt{2}}(\ket{000}+\ket{111})$.}\label{3qubitsgraph} % vector (100) poly x^3+x^2+1
\end{figure}

\begin{figure}[h!]
\includegraphics[width=8.2cm]{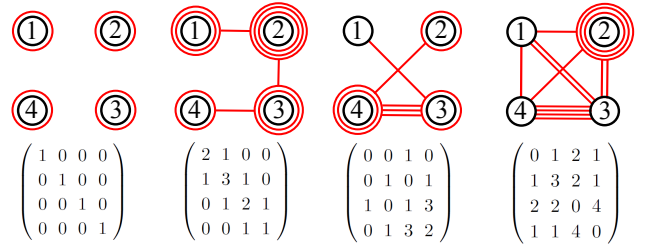}
\caption{(Color online) Fundamental graphs and corresponding adjacency matrices of a complete set of MUBs for four qupits $p=5$ defined by the tridiagonal matrix $Q$, as defined in Eq.~(\ref{tridiag}), with diagonal $\vec{d}=(2,3,2,1)$. A complete set of $5^4$ graphs is obtained through all possible linear combinations of the above adjacency matrices over $\mathbb{Z}_5$.} % vector (2321) poly x^4+2x^3+3x+3
\label{4quditbasis}
\end{figure}

\begin{figure}[h!]
\includegraphics[height=2.5cm]{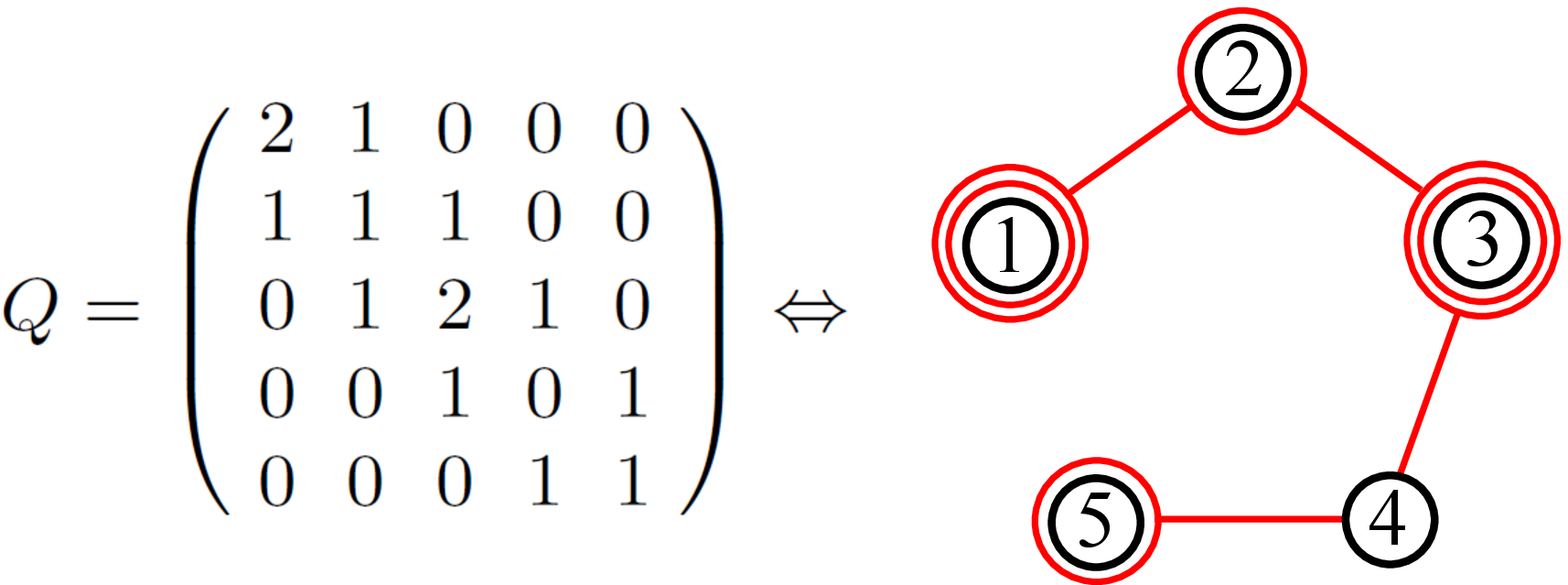}
\caption{(Color online) Example of a primitive graph corresponding to a complete set of MUBs for five qutrits (i.e. $n=5$ and $p=3$) with $\vec{d}=(2,1,2,0,1)$. If the diagonal of the matrix $Q$, as defined in Eq.~(\ref{tridiag}), is chosen such that its characteristic polynomial $\mathrm{char}(Q)$ is a primitive polynomial, then the set of matrix powers $\{ Q^i \}_{i=0}^{p^n-2}$ together with the zero matrix $\mathds{O}_n$ describe a complete set of MUBs.}
% vector (21201) poly x^5+2x+1
\label{5qutritprim}
\end{figure}

Before explaining in detail the construction let us analyze how the corresponding complete set of MUBs looks like. Suppose that $Q$ is a symmetric $n\times n$ matrix such that $S=\{\sum_{i=0}^{n-1} a_i Q^i, \vec{a}=(a_0,\ldots,a_{n-1})\in \mathbbm{Z}_p^n\}$ over $\mathbb{Z}_p$ represents $\F_{p^n}$. Each of the $p^n$ matrices $\sum_{i=0}^{n-1} a_{i} Q^{i}$, with $a_i \in \mathbb{Z}_p$ is an adjacency matrix. According to the discussion above, the corresponding complete set of MUBs is then given by $p^n$ graph--state bases Eq.~(\ref{gengraphstate}), together with the computational basis $\mathcal{B}_C$. Let us now call the graphs corresponding to the $n$ adjacency matrices $\mathcal{F}=\{ Q^0, \ldots, Q^{n-1} \}$ (which constitute a basis of $\F_{p^n}$), \emph{fundamental graphs}. The fact that the adjacency matrices of all the other graphs is just a linear combination of the ones corresponding to the fundamental graphs is also nicely reflected in the corresponding graphs (see Figs.~\ref{2qutritsgraph}, \ref{3qubitsgraph}, and \ref{4quditbasis}). Consider for instance the case $p=2$ and $n=3$. In Fig.~\ref{3qubitsgraph} a complete set of MUBs is depicted. The first three graphs are the fundamental graphs. All the other graphs can be easily read off from those three graphs. For instance, the fourth graph, which corresponds to $Q^0+Q^2$, is obtained by adding all the edges and self--loops modulo $2$ of the graphs corresponding to $Q^0$ and $Q^2$. Similarly all other graphs can be obtained. Thus, it is only necessary to draw the graph of the $n$ fundamental graphs in order to present the complete set of $p^n$ MUBs. As mentioned before, a single matrix $Q$ is required to encode the complete set of MUBs. Likewise, a graph that corresponds to a matrix $Q$ whose characteristic polynomial is primitive, which we call \emph{primitive graph} in the following, encodes the corresponding complete set of MUBs. In Fig.~\ref{5qutritprim} we depict a primitive graph for the case of five qutrits. Whereas the complete set of MUBs can be easily constructed given a primitive graph, the corresponding graphs cannot be easily read off the primitive graph, since they are obtained via matrix multiplication. Note however, that using the presented graph--state formalism in combination with a symmetric matrix $Q$ whose characteristic polynomial is irreducible, it is possible to encode complete sets of MUBs in an extraordinarily compact way. Note further that the matrix $Q$ may also be used to construct a maximally commuting bases as required for the construction presented in \cite{Bandyopadhyay} and discussed in Sec.~\ref{constrbandyo}.

\subsection{Construction via symmetrized companion matrices} \label{symmcomp}
Now, let us discuss how one can find a symmetric matrix $Q$ whose characteristic polynomial is irreducible. We begin with the matrix representation of $\F_{p^n}$ as introduced in Ref.~\cite{Lidl}. The companion matrix $C$ of a monic polynomial $f(x)=x^n+c_{n-1}x^{n-1}+\ldots+c_1 x +c_0$ is defined as the $n \times n$ matrix
\begin{align} \label{companionmatrix}
C=\left(
  \begin{array}{ccccc}
    0  & 1  &   &   &   \\
     & 0  & \ddots  &   &   \\
      &  & \ddots  & \ddots  &  \\
      &  &  &  0 &  1\\
    -c_{0}  & -c_{1}  & \ldots & \ldots & -c_{n-1} \\
  \end{array}
\right) \ .
\end{align}
It is straightforward to show that the characteristic polynomial $\mathrm{char}(C)=\det{(x \mathbbm{1} -C)}$ of the companion matrix equals $f(x)$. Consequently, if $C$ is the companion matrix of a monic irreducible polynomial $f(x)$ of degree $n$ over $\mathbbm{Z}_p$, then the $p^n$ polynomials $a_{n-1} C^{n-1} + \ldots + a_{1} C + a_{0} \mathbbm{1}$ of degree less than $n$ with coefficients $a_k \in \mathbbm{Z}_p$ yield a matrix representation of $\F_{p^n}$, where $\{C^0,\ldots,C^{n-1}\}$ constitutes a basis of $\F_{p^n}$. Hence, given an irreducible polynomial of degree $n$ it is straightforward to construct this matrix representation. However, this representation is not symmetric.

We will now determine a similarity transformation, $P$ (leaving the characteristic polynomial unchanged), such that the companion matrix, $C$, is transformed into a symmetric matrix $Q$. In fact, one can show that any $n \times n$ matrix whose characteristic polynomial is irreducible is similar to the companion matrix \cite{Perlis}. In this way, the whole representation becomes symmetric, since any power of a symmetric matrix and the sum of symmetric matrices is again symmetric. Thus, our aim now is to find an invertible matrix $P$ such that $Q=P C P^{-1}$ satisfies $Q=Q^T$. The existence of such a similarity transformation has already been proven in \cite{Brawley,AAlbert} for any companion matrix of an irreducible polynomial. Hence, the existence of the desired symmetric matrix representation of $\F_{p^n}$ is guaranteed for all $p$ and $n$. Here, we briefly summarize this observation and show how to systematically find $P$, and therefore $Q$, for any $C$ being associated to an irreducible polynomial. Note that an implementation of the following algorithm for Mathematica$\textsuperscript{\textregistered}$ is available online in the Wolfram Library Archive \cite{MathCode}.

First, notice that the requirement $P C P^{-1} = (P C P^{-1})^T$ can straightforwardly be rewritten as $CB=BC^T$ where $B$ is of the form $B=P^{-1}{P^{-1}}^T$. Consequently, finding $P$ can be divided into two steps: First, determine a symmetric invertible matrix $B$ such that $CB=BC^T$. Second, specify a factorization of the form $B=P^{-1}{P^{-1}}^T$. The second step, is equivalent to finding an invertible matrix, $P$, such that $PBP^T=\mathbbm{1}_n$. In order to present a systematic method achieving that, we consider again the two cases, $p=2$ [case (i)] and $p\geq 3$ [case (ii)] separately.

\emph{Case (i) ---} Consider the case where $p=2$. Here, as shown in Ref.~\cite{Brawley}, the symmetric matrix
\begin{align} \label{Amatrix2}
B=\left(
    \begin{array}{cccc}
      1 & 0 & \cdots & 0 \\
      0 &  &  & b_1 \\
      \vdots &  & \iddots & \vdots \\
      0 & b_1 & \cdots & b_{n-1} \\
    \end{array}
  \right) \ ,
\end{align}
where the coefficients $b_k$ are defined via the coefficients $c_k$ of the monic polynomial $f(x)$ of degree $n$ as
\begin{align}
b_1&=c_0 \ , \\
b_i&=\sum_{k=1}^{i-1} c_{n-i+k} b_k \ ,
\end{align}
satisfies the condition $CB=BC^T$. Now, it remains to diagonalize $B$ through a congruence transformation, i.e. to find a matrix $P$ such that $PBP^T=\mathbbm{1}_n$. In Appendix~\ref{congruencep2}, we show how $P$ can be computed using the following toolbox of operations
\begin{align} \label{toolbox1}
  \Pi_{i,j}&=\left(
            \begin{array}{cc}
              0 & 1 \\
              1 & 0 \\
            \end{array}
          \right)_{i,j}&  \Pi_{i,j}^{-1}&=\Pi_{i,j} \nonumber \\
  \Lambda_{i,j}&=\left(
            \begin{array}{cc}
              1 & 0 \\
              1 & 1 \\
            \end{array}
          \right)_{i,j}&  \Lambda_{i,j}^{-1}&=\Lambda_{i,j} \\
  \Omega_{i,j,k}&=\left(
\begin{array}{ccc}
 1 & 1 & 0 \\
 1 & 0 & 1 \\
 1 & 1 & 1 \\
\end{array}
\right)_{i,j,k}&  \Omega_{i,j,k}^{-1}&=\left(
\begin{array}{ccc}
 1 & 1 & 1 \\
 0 & 1 & 1 \\
 1 & 0 & 1 \\
\end{array}
\right)_{i,j,k}
 \ . \nonumber
\end{align}
Here, each matrix $\Pi_{i,j}$, $\Lambda_{i,j}$ or $\Omega_{i,j,k}$ is to be read as an $n \times n$ matrix that affects the rows and columns $i,j,k$ while all other rows/columns remain unchanged (i.e. identity on the rest), e.g. for $n=4$ the matrix $\Lambda_{2,4}$ reads
\begin{align}
\Lambda_{2,4}=\left(
  \begin{array}{cccc}
    1 & 0 & 0 & 0 \\
    0 & 1 & 0 & 0 \\
    0 & 0 & 1 & 0 \\
    0 & 1 & 0 & 1 \\
  \end{array}
\right) \ ,
\end{align}
with the identity on the rows/columns $1$ and $3$.

Specifically, in Appendix~\ref{congruencep2} we show that any non--singular symmetric $n\times n$ matrix $B$ over $\mathbbm{Z}_2$ which has at least one diagonal element equal to $1$ can be transformed into the identity matrix using a sequence of the above operations for congruence transformations. The given proof is constructive and leads to a systematic way to determine the matrix $P$. Since the matrix $B$ given in Eq.~(\ref{Amatrix2}) belongs to this class of matrices, we accomplished the task of finding a similarity transformation $P$, which transforms the companion matrix into the symmetric matrix $PCP^{-1}$ (having the same irreducible characteristic polynomial).

\emph{Case (ii) ---} Consider the case where $p \geq 3$. Again, we seek a matrix $B$ which satisfies $CB=BC^T$ and which is congruent to the identity matrix, i.e. $PBP^T=\mathbbm{1}_n$. As explained above, the matrix $P$ then symmetrizes the companion matrix $C$ via the similarity transformation $PCP^{-1}=Q$. According to Ref.~\cite{Brawley}, for $p \geq 3$ the matrix $B$ can be chosen to be of the form
\begin{align} \label{Amatrixp}
B=g B_0 \ .
\end{align}
Here, $B_0$ is the lower-right triangular and symmetric matrix
\begin{align} \label{B0matrix}
B_0=\left(
      \begin{array}{cccccc}
         &  &  &  &  & 1 \\
         &  &  &  & 1 & b_1 \\
         &  &  & \iddots & b_1 & b_2 \\
         &  & \iddots & \iddots & \iddots & \vdots \\
         & 1 & b_1 & \iddots &  & b_{n-2} \\
        1 & b_1 & b_2 & \cdots & b_{n-2} & b_{n-1} \\
      \end{array}
    \right) \ ,
\end{align}
where the coefficients $b_k$ are defined via the coefficients $c_k$ of the monic polynomial $f(x)$ of degree $n$ as
\begin{align}
b_0&=1 \ , \\
b_i&=-\sum_{k=0}^{i-1} c_{n-i+k} b_k \ ,
\end{align}
and $g$ is either a constant in $\mathbbm{Z}_p$, or a polynomial in the companion matrix $C$ over $\mathbb{Z}_p$ of degree less than or equal to $n-1$, i.e. $a_{n-1} C^{n-1} + \ldots + a_{1} C + a_{0} \mathbbm{1}$.

Here, $g$ has to be chosen such that $\det ( B )$ is a quadratic residue. A quadratic residue $q \in \mathbbm{Z}_p \backslash \{0\}$ is an element which has a square root in $\mathbbm{Z}_p \backslash \{0\}$, i.e. for $q$ there exists an element $s\in \mathbbm{Z}_p \backslash \{0\}$ such that $q=s^2$. An element $\hat{q} \in \mathbbm{Z}_p \backslash \{0\}$ for which there exists no such element is called a quadratic non--residue, that is $\hat{q} \neq s^2$ holds for all $s\in \mathbbm{Z}_p \backslash \{0\}$.

It can be shown that $B$ is congruent to $\mathbbm{1}_n$ iff $\det ( B )$ is a quadratic residue \cite{finitefields}. In Appendix~\ref{congruencep3}, we give a constructive proof of this fact. There, we make use of the toolbox of operations
\begin{align} \label{toolboxp}
  \Pi_{i,j}&=\left(
            \begin{array}{cc}
              0 & 1 \\
              1 & 0 \\
            \end{array}
          \right)_{i,j}&  \Pi_{i,j}^{-1}&=\Pi_{i,j} \nonumber \\
  \Lambda_{i,j}&=\left(
            \begin{array}{cc}
              1 & 0 \\
              a & 1 \\
            \end{array}
          \right)_{i,j}& \Lambda_{i,j}^{-1}&= \left(
            \begin{array}{cc}
              1 & 0 \\
              -a & 1 \\
            \end{array}
          \right)_{i,j} \\
  \Omega_{i,j}&=\left(
\begin{array}{cc}
 1 & 1 \\
 1 & -1 \\
\end{array}
\right)_{i,j}& \Omega_{i,j}^{-1}&=\left[\frac{p+1}{2} \left(
\begin{array}{cc}
 1 & 1 \\
 1 & -1 \\
\end{array}
\right)\right]_{i,j} \nonumber \\
\Phi_{i,j}&=\left(
\begin{array}{cc}
 1 & b \\
 -b & 1 \\
\end{array}
\right)_{i,j}&  \hspace{0.3cm} \Phi_{i,j}^{-1}&=\left[(1+b^2)^{-1} \left(
\begin{array}{cc}
 1 & -b \\
 b & 1 \\
\end{array}
\right)\right]_{i,j} \nonumber
 \ .
\end{align}
In terms of these operations, one obtains a systematic procedure to determine $P$ for which $PBP^T=\mathbbm{1}_n$, where $B$ is any non--singular symmetric $n \times n$ matrix over $\mathbbm{Z}_p$ whose determinant $\det ( B )$ is a quadratic residue. This procedure is presented in Appendix~\ref{congruencep3}.

Now, it remains to discuss how to choose $g$ such that the determinant of $B$ is a quadratic residue. As can easily be seen \cite{Brawley}, for the matrix $B_0$ we have
\begin{align}
\det(B_0) \, = \,\left\{\begin{array}{lll} \ 1 \hspace{0.5cm} & \mbox{if} \ \ (n \ \mathrm{ mod } \ 4)= 0 \mbox{ or } 1  \\
 \\
		\ -1 & \mbox{if} \ \ (n  \ \mathrm{ mod } \ 4) = 2 \mbox{ or } 3 \ . \end{array} \right.
\end{align}
As $1$ is always a quadratic residue we can choose $g=1$ whenever $(n \ \mathrm{ mod } \ 4)= 0$ or $1$. The same also holds for $(n \ \mathrm{ mod } \ 4)= 2$ or $3$ in case for the given $p$ the element $(-1 \ \mathrm{ mod } \ p) \in \mathbbm{Z}_p$ is a quadratic residue \footnote{From number theory it is known that $-1$ is a quadratic residue iff $(p \ \mathrm{ mod } \ 4)= 1$. This relation is known as \emph{`The First Supplement to Quadratic Reciprocity'} \cite{Kenneth}.}. That is, in these cases we can simply choose $B=B_0$. However, if $(n \ \mathrm{ mod } \ 4)= 2$ or $3$ and furthermore $(-1 \ \mathrm{ mod } \ p) \in \mathbbm{Z}_p$ is not a quadratic residue for the particular $p$ we cannot choose $g=1$. In these cases one might proceed as follows (see also Ref.~\cite{Brawley}). If $(n \ \mathrm{ mod } \ 4)= 3$, the number $n$ is odd and $n-1$ is even. For a constant $g \in \mathbbm{Z}_p \backslash \{0\}$ we obtain $\det(B)=\det(gB_0)=g^n \det(B_0)=g^{n-1} (g\det(B_0))$. As $n-1$ is even the factor $g^{n-1}$ is a quadratic residue. Thus, as a product of two non--residues is a quadratic residue \cite{Kenneth} we simply choose $g=\hat{q}$ to be an arbitrary non--residue $\hat{q} \in \mathbbm{Z}_p \backslash \{0\}$ to achieve that the second factor $g\det(B_0)$ becomes a quadratic residue as well. For the remaining case $(n \ \mathrm{ mod } \ 4)= 2$ this does not work as $n-1$ is odd and for any constant $g \in \mathbbm{Z}_p$ the determinant $\det(g B_0)$ remains a non--residue. Here, however, for $f(x)$ being an irreducible polynomial and $C$ being its associated companion matrix, it was shown in Ref.~\cite{Brawley} that there always exists a matrix $g=a_{n-1} C^{n-1} + \ldots + a_{1} C + a_{0} \mathbbm{1}$, which is a polynomial in $C$ over $\mathbb{Z}_p$, with the property that its determinant, $\det(g)$, is a quadratic non--residue. Using such a matrix $g$ the determinant $\det(B)= \det(g) \det(B_0)$ is a product of two non--residues which is again a quadratic residue.

In order to circumvent the search for the coefficients $a_i$ of $g=a_{n-1} C^{n-1} + \ldots + a_{1} C + a_{0} \mathbbm{1}$ such that $\det(g)$ is a quadratic non--residue, the simplest way is to directly choose $f(x)$ to be a primitive polynomial. For any primitive polynomial $f(x)$ over $\mathbbm{Z}_p$ and $p\ge3$ it holds that the determinant of the associated companion matrix, which is $\det(C)=(-1)^n c_0$, is a quadratic non--residue. This is because the element $(-1)^n c_0 \in \mathbb{Z}_p$, where $c_0$ is the lowest coefficient of a primitive polynomial over $\mathbbm{Z}_p$ with $p\geq3$, is always a quadratic non--residue (which is a consequence of Theorem~$3.18$ in Ref.~\cite{Lidl}). Consequently, whenever we have $(n \ \mathrm{ mod } \ 4)= 2$ and $(-1 \ \mathrm{ mod } \ p) \in \mathbbm{Z}_p$ is not a quadratic residue, we simply specify the companion matrix $C$ and the matrix $B_0$ of a primitive polynomial $f(x)$ and choose $g=C$. Then the determinant of the matrix $B=g B_0=C B_0$ is a quadratic residue and, therefore, $B$ is congruent to the identity matrix $\mathbbm{1}_n$.

In summary, we have demonstrated here how a symmetric matrix representation of any finite field $\F_{p^n}$ can be found. That is, we showed how the companion matrix $C$ of an irreducible polynomial $f(x)$ may be symmetrized by means of a constructive algorithm. From this symmetrized companion matrix $Q=P C P^{-1}$, we obtain a set of $p^n$ adjacency matrices via the possible linear combinations of the matrix powers $\{Q^i\}_{i=0}^{n-1}$ over $\mathbb{Z}_p$, i.e. the matrices $a_{n-1} Q^{n-1} + \ldots + a_{1} Q + a_{0} \mathbbm{1}$ for the different settings of the $n$--tuple $(a_{0},\ldots,a_{n-1}) \in \mathbb{Z}^n_p$. Thus, a complete set of MUBs for dimension $d=p^n$ can always be encoded in a single matrix $Q \in \mathbb{Z}_p^{n \times n}$. As the matrix $Q$ is symmetric, it is characterized by $n(n+1)/2$ coefficients from $\mathbb{Z}_p$. An example in which this method is used to construct a complete set of MUBs for $d=3^3=27$ can be found in Appendix~\ref{lowdimensions1}.

\subsection{Construction via tridiagonal matrices}
In this section, we give an alternative way to specify a symmetric matrix $Q$ whose characteristic polynomial is irreducible. In particular, we show that the set of adjacency matrices may even be represented by a single $n$--dimensional vector with coefficients in $\mathbb{Z}_p$. This vector corresponds to the diagonal entries of the symmetric $n\times n$ matrix $Q$ as given in Eq.~(\ref{tridiag}). Note that in the graph--state corresponding to this particular adjacency matrix only nearest neighbor interactions occur.

\begin{figure}[h]
\includegraphics[width=3cm]{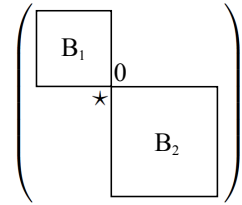}
\caption{The characteristic polynomial $f(x)$ of a tridiagonal matrix which has a zero in the off-diagonal is a product of the characteristic polynomials $f_1(x)$ and $f_2(x)$, of the submatrices $B_1$ and $B_2$. Therefore, the characteristic polynomial $f(x) = f_1(x) \cdot f_2(x)$ is not irreducible.}
\label{blockstructure}
\end{figure}

The following ideas are inspired by the fact that any matrix over the complex numbers is similar to a tridiagonal matrix (see e.g. Ref.~\cite{Horn}). Suppose now the same holds true for matrices over finite fields. In this case it would be sufficient to make an Ansatz for the matrix $Q$ in the tridiagonal form
\begin{align}
\label{tridiaggeneral}
Q=\left(
  \begin{array}{ccccc}
    \star & \star &   &   &   \\
    \star & \star & \star &   &   \\
      & \star & \ddots & \ddots &  \\
      &  & \ddots & \ddots & \star \\
      &   &  & \star & \star \\
  \end{array}
\right) \ ,
\end{align}
where each $\star$ is an arbitrary element of $\mathbb{Z}_p$. Our intention here is to achieve that the characteristic polynomial of $Q$ is irreducible. Consequently, a necessary condition on the tridiagonal matrix $Q$ is that all elements in the sub- and superdiagonal are non--zero, because otherwise the characteristic polynomial would factor into $\mathrm{char}(Q) = \det(x \mathbbm{1} -Q)=f_1(x) \cdot f_2(x)$ according to the block structure ($f_1(x)$ and $f_2(x)$ correspond to the characteristic polynomials of the blocks $B_1$ and $B_2$, respectively; see Fig.~\ref{blockstructure}). In the binary case $\mathbb{Z}_2$, this implies that whenever the characteristic polynomial of a tridiagonal matrix $Q$ is irreducible, it can only be of the form
\begin{align}
\label{tridiag}
Q=\left(
  \begin{array}{ccccc}
    d_1 & 1 &   &   &   \\
    1 & d_2 & 1 &   &   \\
      & 1 & \ddots & \ddots &  \\
      &  & \ddots & \ddots & 1 \\
      &   &  & 1 & d_n \\
  \end{array}
\right) \ ,
\end{align}
because $1$ is the only non--zero element in $\mathbb{Z}_2$. Hence, if there exists a tridiagonal matrix $Q$ over $\mathbb{Z}_2$, whose characteristic polynomial is irreducible, then it is automatically symmetric as desired. Let us focus on this case for the moment. As can easily be seen, if we set $\Delta_0=1$ and $\Delta_{-1}=0$, the characteristic polynomial of the matrix $Q$ given in Eq.~(\ref{tridiag}) satisfies the recursion relation
\begin{align} \label{recursionrel}
\Delta_k=(x-d_{(n+1-k)})\Delta_{k-1}-\Delta_{k-2} \ ,
\end{align}
for $1 \leq k \leq n$, wherein $\Delta_k$ is the characteristic polynomial of the $k\times k$ submatrix defined by the last $k$ components of each row and column of $Q$ (e.g. $\Delta_n$ is simply the characteristic polynomial of $Q$ itself). Note that each polynomial $\Delta_k$ with $1 \leq k \leq n$ has exactly degree $k$. Now, if any irreducible polynomial $f(x)$ is a characteristic polynomial of a particular tridiagonal matrix, it must hold that for $\Delta_{n} \equiv f(x)$ there exists a set of $n$ polynomials $\{\Delta_{k}\}_{k=1}^{n}$, wherein each $\Delta_{k}$ is of degree $k$ and satisfies the recursion relation Eq.~(\ref{recursionrel}) for all $1 \leq k \leq n$. The existence of such a set of polynomials $\{\Delta_{k}\}_{k=1}^{n}$ for any irreducible polynomial with arbitrary degree $n$ was indeed proven in Ref.~\cite{Mesirov}. Hence, for any irreducible polynomial $f(x)$ of degree $n$ over $\mathbb{Z}_2$ there exists a tridiagonal $n \times n$ matrix $Q$ of the form given in Eq.~(\ref{tridiag}), such that $\mathrm{char}(G)=f(x)$.

It now remains to discuss how to find appropriate diagonals for $Q$. The simplest, but very time consuming, method is to straightforwardly compute the characteristic polynomials of $Q$ given in Eq.~(\ref{tridiag}) for different settings of $\vec{d}=(d_1,\ldots,d_n) \in \mathbb{Z}_2^n$ until an irreducible polynomial is found. Another way is to choose an arbitrary irreducible polynomial $f(x)$, and then to tridiagonalize the associated companion matrix $C$. As an example, consider $p(x)=x^3+x+1$ over $\mathbb{Z}_2$ possessing the companion matrix
\begin{align}
C=\left(
    \begin{array}{ccc}
      0 & 1 & 0 \\
      0 & 0 & 1 \\
      1 & 1 & 0 \\
    \end{array}
  \right) \ .
\end{align}
Using
\begin{align}
P=P^{-1}=\left(
    \begin{array}{ccc}
      1 & 0 & 0 \\
      1 & 1 & 0 \\
      0 & 0 & 1 \\
    \end{array}
  \right) \ ,
\end{align}
one obtains
\begin{align}
Q=P C P^{-1}=\left(
    \begin{array}{ccc}
      1 & 1 & 0 \\
      1 & 1 & 1 \\
      0 & 1 & 0 \\
    \end{array}
  \right) \ ,
\end{align}
and thus $\vec{d}=(1,1,0)$. Note that a suitable tridiagonalization algorithm for matrices over $\mathbb{Z}_2$ was introduced in Ref.~\cite{Serra}.

An alternative method to analytically derive $\vec{d}$ is to utilize Newton's identities. For an $n\times n$ matrix $Q$, the following relations between the traces $t_k=\mathrm{tr}(Q^k)$ and the coefficients $c_n$ of the characteristic polynomial $p(x)=x^n+c_{n-1}x^{n-1}+\ldots+c_1 x + c_0$ hold \cite{newtonId}:
\begin{align}
t_1+c_{n-1}=0 \ ,\\
t_k+c_{n-1}t_{k-1}+\ldots+c_{n-k+1}+k c_{n-k}=0 \ ,
\end{align}
where $2\leq k \leq n$. Using these identities enables one to find $\vec{d}$ for a desired irreducible polynomial. Consider again the example $p(x)=x^3+x+1$ over $\mathbb{Z}_2$, i.e. $c_0=1$, $c_1=1$ and $c_2=0$. After elementary simplifications, one finds $t_1=d_1+d_2+d_3$, $t_2=d_1+d_2+d_3$, $t_3=d_2$. Thus we have the relations
\begin{align}
t_1+c_2=d_1+d_2+d_3&=0 \ , \label{equ1} \\
t_2+c_2 t_1 = d_1+d_2+d_3&=0 \ , \label{equ2} \\
t_3+c_2 t_2 + c_1 t_1 + c_0 = d_1+d_3+1 &=0 \label{equ3} \ .
\end{align}
From Eq.~(\ref{equ2}) and Eq.~(\ref{equ3}) it follows that $d_2=1$, and then from Eq.~(\ref{equ1}) that $d_1+d_3=1$. There are two vectors that fulfill $d_2=1$ and $d_1+d_3=1$. Namely, the vectors $\vec{d}=(1,1,0)$ and $\vec{d}=(0,1,1)$. Note that both lead to the same irreducible polynomial $p(x)=x^3+x+1$. Notice that any vector $\vec{d}=(d_1,\ldots,d_n)$ and its reversed counterpart $\vec{d}_r=(d_n,\ldots,d_1)$ always lead to the same characteristic polynomial as their associated matrices (say $Q$ and $Q_r$) are similar.

Finally, note that the tridiagonal matrices discussed here also occur in the context of so-called \emph{one-dimensional linear hybrid cellular automata}. In this regard, a more advanced technique for determining the vector $\vec{d}=(d_1,\ldots,d_n)$ which is based on a quadratic congruence relation was introduced in Ref.~\cite{Cattell}. Using this method, the same authors derived a list of solutions for $n$ up to $300$, which can be found in Ref.~\cite{Cattell300}.

So far, we have just considered the case of tridiagonal matrices over $\mathbb{Z}_2$. Let us now have a brief look at the more general case $\mathbb{Z}_p$. Here, the off-diagonal elements of a tridiagonal matrix whose characteristic polynomial is irreducible can have arbitrary non--zero entries between $1$ and $p-1$. Note that the proof in Ref.~\cite{Mesirov} is restricted to $\mathbb{Z}_2$, and that it is not clear whether there exists for any irreducible polynomial, $f(x)$, a tridiagonal matrix whose characteristic polynomial coincides with $f(x)$. However, by computing the possible settings of the vector $\vec{d}=(d_1,\ldots,d_n)$ with entries in $\mathbb{Z}_p$ in the ansatz given in Eq.~(\ref{tridiag}) for numerous cases, $p$ and $n$, we have made the experience that this form already comprises a variety of irreducible polynomials \footnote{Note that we also observed that in general not all irreducible polynomials can be realized via Eq.~(\ref{tridiag}). Furthermore, numerically we found by the example of $p=3$ and $n=3$ that there also exist irreducible polynomials which are not the characteristic polynomial of any \emph{symmetric} tridiagonal matrix.}. Thus, it could well be that for any $p$ and $n$ there exists a vector $\vec{d}=(d_1,\ldots,d_n) \in \mathbb{Z}^n_p$, defining the diagonal of a tridiagonal matrix $Q$ whose sub- and superdiagonal elements are all $1$, such that the characteristic polynomial of $Q$ is irreducible. An extensive list of solutions, for $p=2,\ldots,7$ and $n=2,\ldots,8$, in terms of vectors $\vec{d}$ describing the diagonal of the tridiagonal matrix Eq.~(\ref{tridiag}) can be found in Appendix~\ref{listprimitive}. An example in which a tridiagonal matrix is used to construct a complete set of MUBs for $d=8$ is given in Appendix~\ref{lowdimensions2}.

\section{Entanglement structures} \label{SECentstrut}
The graph--state formalism is ideally suited to investigate entanglement structures arising in MUBs. That is, all information can readily be obtained simply by looking at the form of the underlying graphs (i.e. the adjacency matrices). For example, for multi-qubits it is well-known (see Ref.~\cite{Hein}) that star graphs and fully connected graphs are local-unitarily (LU) equivalent to the $GHZ$--state $\ket{GHZ}=\ket{0}^{\otimes n} + \ket{1}^{\otimes n}$. In the following, all states with this property shall be referred to as $GHZ$--type states. Thus, e.g. in Fig.~\ref{3qubitsgraph}, we immediately see that $6$ of the $8$ three-qubit MUBs are of $GHZ$--type, while $2$ bases are fully separable. Hence, together with the computational basis (also fully separable), the complete set of MUBs consists of $3$ bases which only contain product states, and $6$ bases which only contain so-called genuinely (or truly) multipartite entangled states \cite{Horodeckireview,Guhnereview}.

This structure generalizes to all three qupit MUBs of arbitrary local dimension $p$. As our construction of the adjacency matrices $\{A_0, \ldots, A_{d-1} \}$ always contains the identity $A_0=\mathbbm{1}_n$ (which is the neutral element of the multiplicative group) and all its multiplicatives over $\mathbbm{Z}_p$, we always obtain $p$ graphs which do not have any edges. That is, all $n$ vertices are isolated in these cases. Together with the computational basis $\mathcal{B}_C$ these graphs give rise to a set of $p+1$ bases whose elements are completely factorized (i.e. separable) with respect to the tripartite Hilbert space $\mathbb{C}^p \otimes \mathbb{C}^p  \otimes \mathbb{C}^p$. For the moment, denote the elements of these $p+1$ bases by $\{ \ket{k_{i}} \otimes \ket{l_{i}} \otimes \ket{m_{i}} \}_{k,l,m=0}^{p-1}$, where the index corresponds to the different bases, i.e. $i \in \{0,\ldots,p\}$. Furthermore, in the tripartite case, any bipartition of the system separates one qupit vs. two qupits, e.g. $\mathcal{H}=\mathcal{H}^{(1)} \otimes \mathcal{H}^{(2,3)}$ where $\mathcal{H}^{(1)}=\mathbb{C}^p$ and $\mathcal{H}^{(2,3)}=\mathbb{C}^{p^2}$. Now, assume there was another graph (besides the $p$ graphs whose adjacency matrices are the multiplicatives of $\mathbbm{1}_n$) that had no connection with respect to the bipartition $(1|23)$. In this case, all elements of the corresponding basis would be separable regarding the splitting $\mathcal{H}^{(1)} \otimes \mathcal{H}^{(2,3)}$ of the Hilbert space. Let us denote these states by $\{\ket{\psi(r)^{(1)}}\otimes \ket{\phi(s)^{(2,3)}}\}$, where $r \in \{0,\ldots,p-1\}$ and $s \in \{0,\ldots,p^2-1\}$. For any pair of basis states, $\ket{\psi(r)^{(1)}}\otimes \ket{\phi(s)^{(2,3)}}$ and $\ket{k_{i}} \otimes \ket{l_{i}} \otimes \ket{m_{i}}$, the mutual overlap factors into $H_3=|\braket{\psi(r)^{(1)}}{k_i}|^2 |\braket{\phi(s)^{(2,3)}}{l_i}\otimes \ket{m_i}|^2=H^{(1)} \cdot H^{(2,3)}$. According to the MU assumption and our construction, we must have $H^{(1)}=1/p$ and $H^{(2,3)}=1/p^2$, such that the overall overlap is $H_3=1/p^3$. This, however, is not possible, as on the Hilbert space $\mathcal{H}^{(1)}=\mathbb{C}^p$, the basis $\{\ket{\psi(r)^{(1)}}\}_{r=0}^{p-1}$ cannot be MU with respect to all $p+1$ bases $\mathcal{B}_i^{(1)}=\{ \ket{k_{i}}\}_{k=0}^{p-1}$, because if this was true we would have found $p+2$ MUBs for a $p$--dimensional Hilbert space; which is of course impossible. Thus, this assumption leads to a contradiction. Consequently, in all the $p^3-p$ remaining graphs each vertex must at least have one outgoing edge to another vertex. Thus, from our construction it follows that for a tripartite qupit system, there always exists a complete set of MUBs consisting of $p+1$ bases whose elements are product states (i.e. fully separable), and $p^3-p$ bases whose elements are entangled with respect to all bipartitions (i.e. genuinely multipartite entangled).

\begin{figure}[h!]
\includegraphics[width=7.5cm]{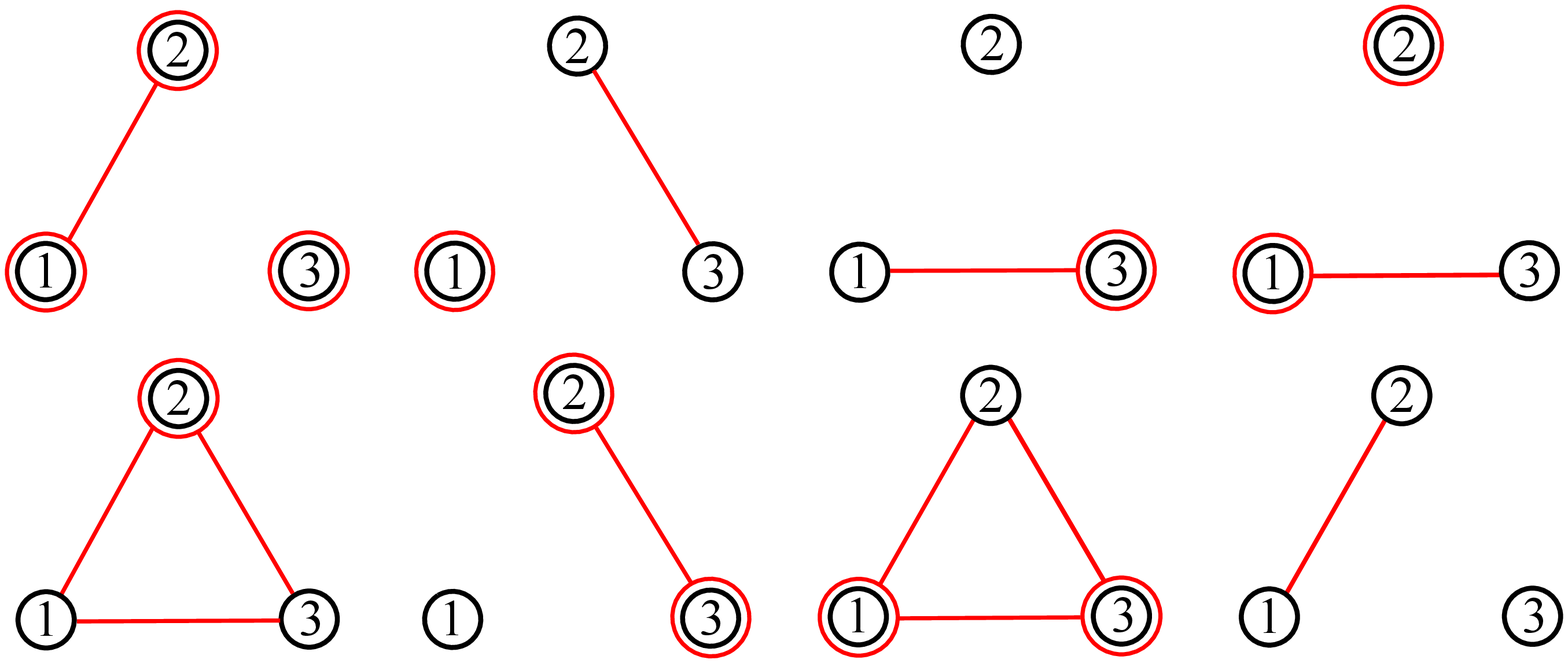}
\caption{(Color online) A complete set of graph--state MUBs for three qubits, which are the MUBs from Fig.~\ref{3qubitsgraph} under the collective action of the controlled phase $U_{1,2}$. Not all vertices of the graphs $1-4$, $6$ and $8$ are connected by an edge. Consequently, the corresponding bases are biseparable. Whereas, the graphs $5$ and $7$ are fully connected, and thus the corresponding graph--states are of $GHZ$--type.}\label{3qubitsmod}
\end{figure}

Besides this, one can also use the graph--state formalism to graphically analyze the action of entangling operations on complete sets of MUBs. As any unitary $U$ which is collectively applied to a set of bases $\{ \mathcal{B}_0, \mathcal{B}_1, \ldots \} \rightarrow \{ U \mathcal{B}_0, U \mathcal{B}_1, \ldots \}$ leaves the scalar product between pairs of basis vectors invariant, i.e. $|\bra{i_k} U^{\dagger} U \ket{j_l}|^2=|\braket{i_k}{j_l}|^2$, it is clear that MU is invariant under such transformations. Using the graph--state formalism, one can illustrate how the entanglement structure changes under certain unitaries. For example, if a controlled entangling operation $U_{1,2}$ is applied between the first two particles on the MUBs in Fig.~\ref{3qubitsgraph}, a connection between the vertices $1$ and $2$ is added $(\mathrm{mod} \ 2)$ and the graphs change to the ones given in Fig.~\ref{3qubitsmod}. Now, the complete set consists of $6$ biseparable bases, $2$ $GHZ$--type bases (fully connected graphs), and $1$ completely separable basis (the computational basis, which is clearly unaltered under $U_{1,2}$ as it only produces global phases in this case).

In general, applying an arbitrary unitary phase gate from the set $\{ U_{i,j} \}$ to a set of graph--state bases $\{ \mathcal{B}_0, \mathcal{B}_1, \ldots \} \rightarrow \{ U_{i,j} \mathcal{B}_0, U_{i,j} \mathcal{B}_1, \ldots \}$ is equivalent to increasing the entries $(i,j)$ and $(j,i)$ of all adjacency matrices $\{ A_0 , A_1, \ldots \}$ by one $( \mathrm{mod} \ p)$. In fact, to any set of $n \times n$ adjacency matrices $\{ A_0 , A_1, \ldots \}$ over $\mathbb{Z}_p$ which satisfy the MU condition $\det (A_r - A_s) \neq 0 \ (\mathrm{mod} \  p), \ \forall r \neq s $ from Lemma \ref{MUBcondition}, we are free to add \emph{any} symmetric matrix $M$ over $\mathbb{Z}_p$, since $\det (A_r +M - (A_s+M))= \det (A_r - A_s)$. Note that each matrix $M$ which has non--zero off-diagonal elements can alter the entanglement properties of a graph--state. Thus, the entanglement structure of a set of MUBs can change as well. Note further that the new set of adjacency matrices $S'=\{ A_0 +M  , A_1 +M, \ldots \}$ does not necessarily constitute a matrix representation of a finite field \footnote{In particular, a non--zero matrix $A_i +M$ is not necessarily invertible.}, as is the case for the original set $S=\{ A_0, A_1, \ldots \}$ using our construction.

In the context of collective unitaries on complete sets of MUBs, i.e. $\{ U \mathcal{B}_1, U \mathcal{B}_2, \ldots \}$, it may also be interesting to apply more general $m$--body phase gates. Here, the resulting states may no longer be graph--states, but belong to the class of LME states \cite{LME}. This gives a new perspective on other constructions such as the one by Alltop \cite{MUBdesigns}, which for $p \geq 5$ was shown to be equivalent to the construction by Wootters and Fields up to a permutation of the vector components \cite{Godsil}. Note that for LME states, such a permutation can always be rephrased in terms of general phase gates \cite{LME}.

\section{MUBs and $2$--designs} \label{SECDesign}
The graph--state formalism also makes it possible to illustrate the $2$--design property of MUBs. A finite set of vectors $D_t= \{ \ket{\psi_i} \}$ in $\mathcal{H}=\mathbb{C}^d$ is called a \emph{complex projective $t$--design} if it holds that
\begin{align}
\int_{\mathcal{H}} |\braket{\phi}{\psi}|^{2k} d\psi = \frac{1}{|D_t|} \sum_{\ket{\psi_i} \in D_t} |\braket{\phi}{\psi_i}|^{2k} \ ,
\end{align}
for all $k \in \{0,1,\ldots,t \}$, and any $\ket{\phi} \in \mathcal{H}$, where $d \psi$ is a unitarily invariant and normalized measure \cite{MUBdesigns,Grossdesign,Haarmeasure}. In other words, $t$--designs enable to compute uniformly weighted integrals over the Hilbert space $\mathcal{H}=\mathbb{C}^d$, where the integrand is a polynomial in $|\braket{\phi}{\psi}|^{2}$ of degree at most $t$, by averaging over a finite set of vectors $D_t= \{ \ket{\psi_i} \}$. In Ref.~\cite{MUBdesigns}, it was shown that the union of the basis vectors of complete sets of MUBs constitute such a design. Namely, mutually unbiased bases are complex projective $2$--designs.

Here, we want to illustrate that this fact is also reflected in the structure of the associated graphs. In Ref.~\cite{Page}, it was found that the average purity of a reduced density matrix on a bipartite Hilbert space $\mathcal{H}=\mathcal{H}_X \otimes \mathcal{H}_Y=\mathbb{C}^{d_X} \otimes \mathbb{C}^{d_Y}$ over all pure states is given by
\begin{align}
\label{avpurity}
 \langle  \mathrm{tr}(\rho_X^2) \rangle \equiv \int_{\mathcal{H}} \mathrm{tr}(\rho_X^2) d\psi= \frac{d_X+d_Y}{d_X d_Y+1} \ .
\end{align}
where $\rho_X=\mathrm{tr}_Y(\rho)$ is the reduced density matrix of $\rho= \ket{\psi}\bra{\psi}$, and the average is taken with respect to $d \psi$ as previously described. The integrand $\mathrm{tr}(\rho_X^2)$ of this expression contains absolute squares of vector components up to the power $k=2$ \cite{Grossdesign,Haarmeasure}, and can thus be computed by means of a complex projective $2$--design $D_2$, i.e.
\begin{align} \label{avpureplace}
 \langle  \mathrm{tr}(\rho_X^2) \rangle = \frac{1}{|D_2|} \sum_{\ket{\psi_i}\in D_2} \mathrm{tr}({\rho_i}_X^2) \ ,
\end{align}
using a finite number of states $\rho_i=\ket{\psi_i}\bra{\psi_i}$ from the set $D_2=\{ \ket{\psi_i} \}$. As mentioned above, a possible choice for the set $D_2$ is the set of all basis vectors of an arbitrary complete set of MUBs. Now, recall that in our framework all vectors $\{ \ket{G_r(m_1,\ldots,m_n)} \}$ within one graph--state basis $\mathcal{B}_r$ are LU equivalent. Thus, in order to evaluate $\langle  \mathrm{tr}(\rho_X^2) \rangle$ from Eq.~(\ref{avpureplace}), we only need to average over $d+1$ vectors, i.e. one element per basis. Second, the purity $\mathrm{tr}(\rho_X^2)$ of the reduced state $\rho_X$ of a graph--state with respect to an arbitrarily chosen bipartition $(X|Y)$ of the Hilbert space, can directly be derived from the corresponding adjacency matrix, $A$ \cite{Hein}. In fact, it suffices to determine the rank of the connectivity submatrix $\Gamma^{(X|Y)}$, which for a given adjacency matrix $A$ is defined as
\begin{align}
A=\left(
  \begin{array}{c}
    \begin{tabular}{| c |   c  |}
      \hline
      & \\
       X  & \hspace{0.1cm} $\Gamma^{(X|Y)}$ \hspace{0.1cm}   \\
     & \\  \hline
     & \\
     \ ${\Gamma^{(X|Y)}}^T$ \ &  Y   \\
       & \\
      \hline
    \end{tabular}
     \\
  \end{array}
\right) \ ,
\end{align}
wherein the blocks $X$ and $Y$ correspond to the bipartition of the Hilbert space $\mathcal{H}=\mathcal{H}_X \otimes \mathcal{H}_Y$, with $\mathcal{H}_X$ $(\mathcal{H}_Y)$ being the Hilbert space of $n_X$ $(n_Y)$ qupits such that $n_X+n_Y=n$. As shown in Ref.~\cite{Hein}, the purity of a reduced graph--state on $\mathcal{H}_X$ is $\mathrm{tr}(\rho_X^2)=p^{-\mathrm{rank}(\Gamma^{(X|Y)})}$, wherein $\mathrm{rank}(\Gamma^{(X|Y)})$ is computed over $\mathbb{Z}_p$. Thus, given the adjacency matrices $\{A_0, \ldots, A_{d-1} \}$ of $d$ graphs that correspond to a complete set of MUBs, we have that the average purity from Eq.~(\ref{avpurity}) satisfies the relation
\begin{align} \label{mubpurity}
\langle \mathrm{tr}(\rho_X^2) \rangle= \frac{1}{d+1} \left(1+ \sum_{i=0}^{d-1} p^{-\mathrm{rank}(\Gamma^{(X|Y)}_i)}\right) \ ,
\end{align}
wherein $\{\Gamma^{(X|Y)}_1, \ldots, \Gamma^{(X|Y)}_d\}$ are the connectivity submatrices of $\{A_0, \ldots, A_{d-1} \}$. Herein, the first term in the bracket, i.e. the number $1$, stems from the computational basis which is separable and hence $\mathrm{tr}(\rho_X^2)=1$; whereas the second term follows from $\mathrm{tr}({\rho_i}_X^2)=p^{-\mathrm{rank}(\Gamma^{(X|Y)}_i)}$ for each graph--state $\ket{G_i}$. Note that the combination of Eq.~(\ref{avpurity}) and Eq.~(\ref{mubpurity}) yields the necessary condition $1+ \sum_{i=0}^{d-1} p^{-\mathrm{rank}(\Gamma^{(X|Y)}_i)}=d_X+d_Y$ on the adjacency matrices $\{A_0, \ldots, A_{d-1} \}$ for any complete set of graph--state MUBs.

Let us illustrate this connection by the example of a $3$--qubit system using the MUBs shown in Fig.~\ref{3qubitsgraph} and Fig.~\ref{3qubitsmod}. Consider an arbitrary bipartition of the system, say $(1 | 23 )$. Here, for each graph $G_i$ which has a connection with respect to the bipartition $(1 | 23 )$ the $1 \times 2$--dimensional connectivity matrix $\Gamma^{(1 | 23 )}_i$ has rank $1$ and hence the corresponding graph--state contributes a purity of $\mathrm{tr}({\rho_i}_X^2)=p^{-1}=1/2$. On the other hand, if for a graph there is no connection between $(1 | 23 )$ then the rank of $\Gamma^{(1 | 23 )}_i$ is $0$ and therefore, in these cases, $\mathrm{tr}({\rho_i}_X^2)=p^{-0}=1$. In both figures, Fig.~\ref{3qubitsgraph} and Fig.~\ref{3qubitsmod}, we see that six of the eight graphs have connections with respect to the bipartition $(1 | 23 )$, while two of them do not. Thus, using Eq.~(\ref{mubpurity}) we obtain $\langle \mathrm{tr}(\rho_X^2) \rangle= \frac{1}{9} (1 + 2\times1 + 6 \times \frac{1}{2})=\frac{6}{9}$ in agreement with Eq.~(\ref{avpurity}).

This result can be generalized to any bipartition of an arbitrary three-qupit system. As explained in the previous Sec.~\ref{SECentstrut}, for a tripartite qupit system there always exists a complete set of $p^3$ graphs, of which $p$ graphs are completely disconnected (i.e. all $n$ vertices are isolated), and $p^3-p$ graphs have no isolated vertices. For such a complete set we obtain the following. The off-diagonal elements of the completely disconnected graphs are all zero, and thus for them, the rank of the connectivity matrix $\Gamma^{(1 | 23 )}_i$ is zero. Consequently, $\mathrm{tr}({\rho_i}_X^2)=p^{-0}=1$ for these graphs. On the other hand, for the $p^3-p$ graphs with no isolated vertices we have $\mathrm{tr}({\rho_i}_X^2)=p^{-1}$, as the rank of the $1 \times 2$ connectivity matrix $\Gamma^{(1 | 23 )}_i$ is $1$ if a vertex has at least one outgoing connection. Consequently, using Eq.~(\ref{mubpurity}) we obtain $\langle \mathrm{tr}(\rho_X^2) \rangle= \frac{1}{p^3+1}(1 + p \times1 + (p^3-p) \times \frac{1}{p})=\frac{p+p^2}{p^3+1}$. This result is again consistent with Eq.~(\ref{avpurity}).
\section{Implementation} \label{SECImplement}
Several schemes of quantum key distribution \cite{Brusssecret,Cerfsecret}, state tomography \cite{wootters,Adamson}, and entanglement detection \cite{mubdetection} rely on measuring observables whose eigenbases are mutually unbiased. In this section, we discuss how such measurements may be experimentally realized using the MUBs presented in this paper.

\begin{figure}[h!]
\includegraphics[height=3cm]{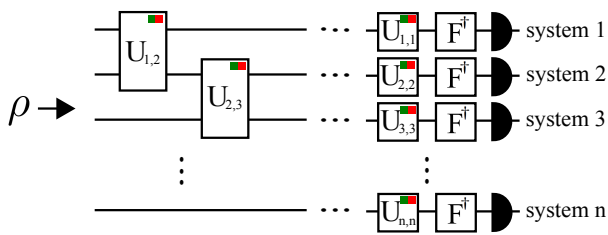}
\caption{(Color online) Implementation of MUBs. This illustration shows the basic circuit that establishes a measurement in a generalized graph--state basis. First, the state $\rho$ undergoes a sequence of two-body controlled phase gates $U_{i,j}$, and local phase gates $U_{i,i}$. Depending on which basis is to be realized, specific gates have to be switched on/off (symbolized by the green/red switches), or applied several times. Finally, a joint measurement is performed in the Fourier basis. This is equivalent to locally applying $F^{\dagger}$ and measuring in the computational basis (black semicircles).} \label{circuit}
\end{figure}

In experiments, we are generally interested in the probabilities $P_k(i)=|\bra{i_k} \rho \ket{i_k} |^2$
of obtaining the outcome $i \in \{0,\ldots,d-1\}$ in the measurement setting $k$, for a system in the state $\rho$. If the setting $k$ corresponds to a basis from a complete set of MUBs from our construction, then each measurement outcome $i$ is related to a particular configuration of the $n$--tuple $(m_1,\ldots,m_n) \in \mathbb{Z}_p^n$, which is related to either a state of the computational basis $\ket{m_1}\otimes \ldots \otimes \ket{m_n}$ or a graph--state $\ket{G(m_1,\ldots,m_n)}$. Thus, in the present case the probabilities are either of the form $P_C(m_1,\ldots,m_n)=|\bra{m_1} \cdots \bra{m_n}\rho \ket{m_1} \cdots  \ket{m_n} |^2$, or $P_k(m_1,\ldots,m_n)=| \bra{G_k(m_1,\ldots,m_n)} \rho \ket{G_k(m_1,\ldots,m_n)} |^2$. If the underlying physical system is indeed a composite $n$--body qupit system, it is generally required to decompose a measurement into experimentally accessible joint probabilities. For a measurement in the computational basis $\mathcal{B}_C$ this is directly the case. On the other hand, for a measurement in a graph--state basis $\mathcal{B}_G$ we have that $\ket{G(m_1,\ldots,m_n)}= U_G F^{\otimes n} \left(\bigotimes_{i=1}^{n} \ket{m_i} \right)$, where $U_G=\prod_{i \leq j} U_{i,j}^{A_{i,j}}$ is the unitary operator from Eq.~(\ref{graphstateconstr}) defining the graph--state, and $F= \frac{1}{\sqrt{p}} \sum_{i,j=0}^{p-1} \omega_{p}^{j i} \ket{i} \bra{j}$ is a local Fourier transform \cite{quditgraphs2}. Therefore, $P_k(m_1,\ldots,m_n)$ is the joint probability of the local measurement outcomes $m_1,\ldots,m_n$ in the Fourier basis with the system being in the state $U_G^{\dagger} \rho U_G$. Thus in summary, one procedure to specify probabilities in a graph--state basis is:
\begin{enumerate}
  \item Let the state undergo the unitary transformation $ \rho \rightarrow U_G^{\dagger} \rho U_G$ where $U_G^{\dagger}= \prod_{i \leq j} U_{i,j}^{-A_{i,j}}$.
  \item Measure the joint probabilities $P(m_1,\ldots,m_n)$ of the state $U_G^{\dagger} \rho U_G$ in the (local) Fourier basis.
\end{enumerate}
This constitutes an experimentally friendly implementation of measurements in MUBs, since it can be realized with only three fundamental operations. The local Fourier transform $F$, the local phase gate $U_{i,i}$, and the two-body controlled phase operation $U_{i,j}$. Accordingly, to experimentally measure in this complete sets of MUBs, only few physical devices are needed; which are then adjusted according to the desired measurement setting. In particular, for a multi-qubit system $d=2^n$, the construction only requires standard gates from quantum computing. Namely, the Hadamard gate $H=F=\frac{1}{\sqrt{2}}(\ket{0}\bra{0}+\ket{1}\bra{0}+\ket{0}\bra{1}-\ket{1}\bra{1})$, the $\pi/4$--phase shift gate $R_{\pi/4}=U_{i,i}=\ket{0}\bra{0}+\mathbbm{i}\ket{1}\bra{1}$, and the controlled-$Z$ gate $CZ=U_{i,j}=\ket{00}\bra{00}+\ket{01}\bra{01}+\ket{10}\bra{10}-\ket{11}\bra{11}$. Note that in order to realize $d+1$ MUBs one may always use the same experimental setup (see Fig.~\ref{circuit}), but with different elements of the circuit switched on or off, depending on which setting is to be generated.

\section{Summary} \label{SECConclusion}
In this paper, we developed a graph--state formalism for the construction of mutually unbiased bases in prime power dimensions, $d=p^n$. We showed that a pair of graph--state bases are mutually unbiased if the difference between the corresponding adjacency matrices has a non--zero determinant over $\mathbb{Z}_p$. In order to construct complete sets of MUBs, we used the theory of finite fields. Namely, we showed that the required condition is automatically fulfilled in case the set of adjacency matrices represents a finite field $\mathbb{F}_{p^n}$. We presented an explicit construction yielding a symmetric matrix representation for any finite field of arbitrary order. Here, we showed that a complete set of adjacency matrices can be generated from a single $n \times n$ matrix, and gave a constructive algorithm to derive this matrix. Moreover, we discussed that in general it is sufficient to specify a single $n$--dimensional vector to construct a complete set of MUBs. Based on this description, we found that any adjacency matrix is a linear combination of $n$ fundamental adjacency matrices. Besides the fact that the introduced construction of MUBs is comparatively simple and illustrative, we have discussed several advantages of our formalism. For example, our framework yields an experimentally friendly physical implementation in terms of only three fundamental gates. Furthermore, the presented formalism is ideally suited to investigate entanglement structure within sets of MUBs. In this direction, further research may be carried out to better understand the role of entanglement in MUBs. In particular, the condition on the average purity of mutually unbiased basis states that follows from the $2$--design property may be useful to investigate the possible non--existence of complete sets of MUBs for non--prime power dimensions, or to exclude certain classes of constructions of MUBs for those dimensions.\\
\\
\noindent \textbf{Acknowledgments:} This research was funded by the Austrian Science Fund (FWF): Y535-N16.\\
\\
\noindent \textbf{Note added:} After submission of the manuscript to the journal we learned that there is a further paper \cite{Seroussi} which proves the existence of symmetric matrix representations over $\mathbbm{Z}_p$ for all finite fields $\F_{p^n}$. Furthermore, related connections between affine planes over finite fields and MUBs were also established in \cite{Kantor}. We would like to thank Markus Grassl for pointing this out to us.

\appendix

\section{Systematic congruence transformation into identity}
In this appendix, we give a constructive algorithm to reduce a symmetric non--singular $n \times n$ matrix over $\mathbb{Z}_p$ [as given in Eq.~(\ref{Amatrix2}) and Eq.~(\ref{Amatrixp})] to the identity matrix $\mathbbm{1}_n$ by means of a sequence of congruence transformations.
\subsection{Case (i) --- $p=2$} \label{congruencep2}
We show that, using the operations from the toolbox Eq.~(\ref{toolbox1}), any non--singular symmetric $n\times n$ matrix $B$ over $\mathbbm{Z}_2$ which has at least one $1$ on the diagonal can be transformed into the identity matrix. (Note that the matrix from Eq.~(\ref{Amatrix2}) belongs to this class of matrices.) To show this, one can proceed in a straightforward fashion similar to a Gaussian elimination: First, the $(1,1)$ element of the matrix $B$ is made non--zero. Either this is already the case, or we apply $\Pi_{1,j}B\Pi_{1,j}^T=B'$ to permute an arbitrary $1$ on the diagonal at position, say $(j,j)$, to $B'_{1,1}$. In the next step, we perform $\Lambda_{1,j} B' \Lambda_{1,j}^T=B''$ on all entries $j\geq 2$ for which $B'_{j,1}\neq0$. In this way, all entries except the first entry of the first column and row become $0$. Let us denote this matrix by $B^{(1)}$. If $B^{(1)}$ has further non--zero diagonal elements besides the element $(1,1)$, we can do the same for the next column and row. That is, if necessary we perform a permutation $\Pi_{2,j}B^{(1)}\Pi_{2,j}^T={B^{(1)}}'$ with $j\geq2$ to make the $(2,2)$ element non--zero, and then the elimination $\Lambda_{2,j} {B^{(1)}}' \Lambda_{2,j}^T={B^{(1)}}''$ on all entries $j\geq 3$ for which ${B^{(1)}}'_{j,2}\neq0$. As the applied operations leave the first column and row invariant, we obtain a matrix, say $B^{(2)}$, whose only non--zero elements in the first two columns and rows are the two diagonal elements $(1,1)$ and $(2,2)$. This elimination is repeated for the next columns/rows as long as after each step the new matrix $B^{(k)}$, which acts like the identity on the first $k$ columns/rows, has a non--zero diagonal element $B^{(k)}_{m,m}$ with $m>k$. Either we directly obtain the $n \times n$ identity matrix in this way, or we arrive at a matrix whose diagonal elements $B^{(k)}_{m,m}$ with $m>k$ are all zero. In this case we proceed as follows. According to our assumption, $B^{(k)}$ is non--singular. Therefore, there must exist at least one non--zero entry in the $(k+1)$'th column of $B^{(k)}$. In order to keep track of the order of our transformation, we want this to be the element $(k+2,k+1)$, and therefore, if necessary, we permute it to this position by applying $\Pi_{k+2,j}B^{(k)}\Pi_{k+2,j}={B^{(k)}}'$, where $j$ corresponds to a non--zero element of column $k+1$. Next, we can exploit (see e.g. Ref.~\cite{AAlbert}) that
\begin{align}
\left(
\begin{array}{ccc}
 1 & 1 & 0 \\
 1 & 0 & 1 \\
 1 & 1 & 1 \\
\end{array}
\right) \left(
                 \begin{array}{ccc}
                   1 & 0 & 0 \\
                   0 & 0 & 1 \\
                   0 & 1 & 0 \\
                 \end{array}
               \right) \left(
\begin{array}{ccc}
 1 & 1 & 0 \\
 1 & 0 & 1 \\
 1 & 1 & 1 \\
\end{array}
\right)^T= \left(
                                                   \begin{array}{ccc}
                                                     1 & 0 & 0\\
                                                     0 & 1 & 0 \\
                                                     0 & 0 & 1 \\
                                                   \end{array}
                                                 \right) \ .
\end{align}
Hence, after the congruence transformation $\Omega_{k,k+1,k+2}{B^{(k)}}'\Omega_{k,k+1,k+2}^T={B^{(k)}}''$ the diagonal elements ${B_{k,k}^{(k)''}}$,${B_{k+1,k+1}^{(k)''}}$ and ${B_{k+2,k+2}^{(k)''}}$ are all $1$. Subsequently, we are again able to eliminate all off-diagonal elements of the corresponding columns and rows by performing $\Lambda_{i,j} {B^{(k)}}'' \Lambda_{i,j}^T$ to all $i=k,k+1,k+2$ (in ascending order) for which the corresponding off-diagonal elements ${B_{i,j}^{(k)}}''$ are non--zero. Note that by applying $\Omega_{k,k+1,k+2}$, new non--zero elements may have been introduced to column/row $k$, which have to be eliminated again. In total, we obtain a new matrix $B^{(k+2)}$ which acts like the identity on the first $k'=k+2$ columns/rows. These steps are repeated until we arrive at the overall identity matrix $B^{(n)}=\mathbbm{1}_n$. Note that this procedure always successfully leads to the identity if the given matrix $B$ is non--singular and has at least one diagonal element which is $1$. If this procedure is applied to the matrix from Eq.~(\ref{Amatrix2}), we obtain $P$ which diagonalizes $B$ via congruence transformation $PBP^T=\mathbbm{1}_n$. Consequently, the same matrix $P$ then symmetrizes the associated companion matrix $C$ [Eq.~(\ref{companionmatrix})] via the similarity transformation $PCP^{-1}=Q$.

\subsection{Case (ii) --- $p\geq3$} \label{congruencep3}
We show that, using the operations from the toolbox Eq.~(\ref{toolboxp}), any non--singular symmetric $n\times n$ matrix $B$ over $\mathbbm{Z}_p$, whose determinant is a quadratic residue, can be transformed into the identity matrix via congruence transformations. We diagonalize $B$, similar to the case $p=2$, by eliminating off-diagonal elements column by column. Again, we want to make the $(1,1)$ element non--zero. Let us first assume that there exists a non--zero element on the diagonal. If necessary we may permute a non--zero element $(j,j)$ on the diagonal to $(1,1)$ using the permutation $\Pi_{1,j} B \Pi_{1,j}=B'$. Subsequently, we can eliminate all off-diagonal elements of the first column and row by applying $\Lambda_{1,j} B' \Lambda_{1,j}=B''$ for all $j\geq 2$ for which $B'_{j,1}$ is non--zero. In order to achieve this, the coefficient $a$ in each $\Lambda_{1,j}$ has to be chosen such that $a B'_{1,1}+B'_{j,1}=0$. Now the only non--zero element of the first column and row is the $(1,1)$ element. If possible, i.e. if there are further diagonal elements besides $(1,1)$, we can repeat this for columns $2,3,4,\emph{etc.}$. However, in case there are no non--zero elements on the diagonal besides the ones for which the elimination has already been performed, or in case $B$ had no diagonal elements from the beginning, we cannot make further diagonal elements non--zero using merely the permutations $\Pi_{i,j}$. Nonetheless, for $p\geq3$ any symmetric $2 \times 2$ block with empty diagonal can be diagonalized via
\begin{align}
\left(
\begin{array}{cc}
 1 & 1 \\
 1 & -1 \\
\end{array}
\right) \left(
                 \begin{array}{cc}
                   0 & d \\
                   d & 0 \\
                 \end{array}
               \right) \left(
\begin{array}{cc}
 1 & 1 \\
 1 & -1 \\
\end{array}
\right) ^T= \left(
\begin{array}{cc}
 2d & 0 \\
 0 & -2d \\
\end{array}
\right) \ .
\end{align}
Thus, whenever we cannot make a diagonal element $(k,k)$ non--zero via a permutation, we permute a non--zero off-diagonal element $(j,k)$ to $(k+1,k)$, and then apply $\Omega_{k,k+1}$. As in this way a pair of diagonal elements, namely $(k,k)$ and $(k+1,k+1)$, become non--zero. Subsequently, we can again eliminate the corresponding off-diagonal elements of column $k$ and $k+1$, if necessary. This is repeated until the matrix is diagonal.

After this diagonalization each individual diagonal element can be a quadratic residue, or a quadratic non--residue. However, as we assume that the determinant of the matrix $B$ is a quadratic residue, the number of quadratic non--residues must be even. This follows from the fact that a product of any two quadratic non--residues is a quadratic residue, whereas the product of a non--residue with a residue yields a quadratic non--residue \cite{Kenneth}. Furthermore, for a pair of quadratic non--residues, $\hat{q}_1$ and $\hat{q}_2$, there always exists a quadratic residue $q=s^2$ such that $\hat{q}_1=q \hat{q}_2$ \footnote{This follows trivially from the fact that in $\mathbbm{Z}_p \backslash \{0\}$ there are $(p-1)/2$ quadratic residues, and $(p-1)/2$ quadratic non--residues \cite{Kenneth}.}. We use these facts to proceed as follows. First, for any quadratic residue $q=s^2$ on the diagonal we can use
\begin{align} \label{residuetrafo} \left(
  \begin{array}{ccc}
    \mathbbm{1} &  &  \\
     & s^{-1} &  \\
     &  & \mathbbm{1}\\
  \end{array}
\right) \left(
  \begin{array}{ccc}
    \star &  &  \\
     & q &  \\
     &  & \star \\
  \end{array}
\right) \left(
  \begin{array}{ccc}
    \mathbbm{1} &  &  \\
     & s^{-1} &  \\
     &  & \mathbbm{1} \\
  \end{array}
\right) =
\left(
  \begin{array}{ccc}
    \star &  &  \\
     & 1 &  \\
     &  & \star \\
  \end{array}
\right)\ ,
\end{align}
where each $\star$ denotes an arbitrary entry. Hence, by applying such congruence transformations to all quadratic residues on the diagonal, we can make these entries equal to $1$. Next, using a further diagonal matrix for congruence transformation we can make all non--residues equal, say $\hat{q}$. Recall that the number of quadratic non--residues is even, i.e. they come in pairs. On all pairs, say elements $(i,i)$ and $(j,j),$ we apply the congruence transformation $\Phi_{i,j}$, which yields
\begin{align}
\left(
\begin{array}{cc}
 1 & b \\
 -b & 1 \\
\end{array}
\right) \left(
\begin{array}{cc}
 \hat{q} & 0 \\
 0 & \hat{q} \\
\end{array}
\right) \left(
\begin{array}{cc}
 1 & -b \\
 b & 1 \\
\end{array}
\right) = \left(1+b^2\right)
\left(
\begin{array}{cc}
 \hat{q} & 0 \\
 0 & \hat{q} \\
\end{array}
\right) \ .
\end{align}
If $b$ is chosen such that $1+b^2$ is an arbitrary quadratic non--residue $\hat{q}'$, then the product $(1+b^2)\hat{q}=\hat{q}' \hat{q}$ is a quadratic residue. It is easy to see that this choice is always possible, and hence the diagonal elements become quadratic residues $q$. Subsequently, we can use again a diagonal matrix Eq.~(\ref{residuetrafo}) to make all diagonal element equal to $1$. Thus, overall we successfully obtain the matrix $P$ for which $PBP^T=\mathbbm{1}_n$. If this procedure is applied to the matrix $B$ from Eq.~(\ref{Amatrixp}), we obtain $P$ for which the matrix $Q=PCP^{-1}$ is symmetric; wherein $C$ is the companion matrix Eq.~(\ref{companionmatrix}).

\section{List of tridiagonal solutions} \label{listprimitive}
In this appendix we give a list of vectors $\vec{d}=(d_1,\ldots,d_n)$ for the tridiagonal $n \times n$ matrix $Q$ [see Eq.~(\ref{tridiag})] such that $f(x)=\mathrm{char} (Q)$ is a (monic) irreducible polynomial of degree $n$. Here, $c_0, \ldots, c_{n-1}$ denote the coefficients of the polynomial in the form $f(x)=x^n+c_{n-1}x^{n-1}+\ldots + c_1 x + c_0$. Note that the shown irreducible polynomials $f(x)$ are also primitive. Also see Ref.~\cite{Cattell300} for further solutions of this form for the special case $p=2$ up to $n=300$.
\subsection{Qubits $p=2$}
\flushleft{ $\begin{array}{|cc|cc|}
  \hline
  \multicolumn{4}{|l|}{\hspace{0.4cm}n=2}\\
  \hline
  d_1 & d_2 & c_1 & c_0 \\
  \hline
  1 & 0 & 1 & 1\\
  \hline
\end{array}$\\
\vspace{0.1cm}
$\begin{array}{|ccc|ccc|}
  \hline
  \multicolumn{6}{|l|}{\hspace{0.4cm}n=3}\\
  \hline
  d_1 & d_2 & d_3 & c_2 & c_1 & c_0\\
  \hline
 1 & 1 & 0 & 0 & 1 & 1\\ 1 & 0 & 0 & 1 & 0 & 1 \\
  \hline
\end{array}$ \\
\vspace{0.1cm}
$\begin{array}{|cccc|cccc|}
\hline
  \multicolumn{8}{|l|}{\hspace{0.4cm}n=4}\\
  \hline
  d_1 & d_2 & d_3& d_4  &  c_3 & c_2 & c_1 & c_0 \\
  \hline
 1 & 0 & 1 & 0 & 0 & 0 & 1 & 1\\ 1 & 1 & 0 & 1 & 1 & 0 & 0 & 1 \\
  \hline
\end{array}$ \\
\vspace{0.1cm}
$\begin{array}{|ccccc|ccccc|}
\hline
  \multicolumn{10}{|l|}{\hspace{0.4cm}n=5}\\
  \hline
  d_1 & d_2 & d_3& d_4 & d_5  & c_4 & c_3 & c_2 & c_1 & c_0\\
  \hline
  1 & 1 & 1 & 1 & 0 & 0 & 0 & 1 & 0 & 1\\ 0 & 1 & 1 & 0 & 0 & 0 & 1 & 0 & 0 & 1\\ 1 & 1 & 0 & 0 & 0 & 0 & 1 & 1 & 1 & 1\\ 1 & 0 & 0 & 0 & 0 & 1 & 0 & 1 & 1 & 1 \\
  \hline
\end{array}$ \\
\vspace{0.1cm}
$\begin{array}{|cccccc|cccccc|}
\hline
  \multicolumn{12}{|l|}{\hspace{0.4cm}n=6}\\
  \hline
  d_1 & d_2 & d_3& d_4 & d_5 & d_6 & c_5 & c_4 & c_3 & c_2 & c_1 & c_0\\
  \hline
  0 & 1 & 1 & 0 & 0 & 0 & 0 & 0 & 0 & 0 & 1 & 1\\ 1 & 0 & 1 & 1 & 1 & 0 & 0 & 1 & 1 & 0 & 1 & 1\\ 0 & 1 & 1 & 0 & 1 & 0 & 1 & 0 & 0 & 0 & 0 & 1\\ 1 & 0 & 1 & 0 & 0 & 1 & 1 & 0 & 0 & 1 & 1 & 1 \\
  \hline
\end{array}$ \\
\vspace{0.1cm}
$\begin{array}{|ccccccc|ccccccc|}
\hline
  \multicolumn{14}{|l|}{\hspace{0.4cm}n=7}\\
  \hline
  d_1 & d_2 & d_3& d_4 & d_5 & d_6 & d_7 & c_6 & c_5 & c_4 & c_3 & c_2 & c_1 & c_0\\
  \hline
 1 & 0 & 1 & 1 & 0 & 0 & 1 & 0 & 0 & 0 & 0 & 0 & 1 & 1\\ 0 & 1 & 1 & 1 & 0 & 1 & 0 & 0 & 0 & 0 & 1 & 0 & 0 & 1\\ 1 & 1 & 1 & 0 & 0 & 0 & 1 & 0 & 0 & 0 & 1 & 1 & 1 & 1\\ 1 & 1 & 1 & 0 & 1 & 0 & 0 & 0 & 0 & 1 & 0 & 0 & 0 & 1 \\
  \hline
\end{array}$ \\
\vspace{0.1cm}
$\begin{array}{|cccccccc|cccccccc|}
\hline
  \multicolumn{16}{|l|}{\hspace{0.4cm}n=8}\\
  \hline
  d_1 & d_2 & d_3& d_4 & d_5 & d_6 & d_7 & d_8 & c_7 & c_6 & c_5 & c_4 & c_3 & c_2 & c_1 & c_0\\
  \hline
  0 & 1 & 1 & 0 & 0 & 0 & 0 & 0 & 0 & 0 & 0 & 1 & 1 & 1 & 0 & 1\\ 1 & 1 & 1 & 1 & 1 & 0 & 1 & 0 & 0 & 0 & 1 & 0 & 1 & 0 & 1 & 1\\ 1 & 1 & 1 & 0 & 1 & 1 & 1 & 0 & 0 & 0 & 1 & 0 & 1 & 1 & 0 & 1\\ 0 & 1 & 1 & 0 & 1 & 1 & 0 & 0 & 0 & 1 & 0 & 0 & 1 & 1 & 0 & 1 \\
  \hline
\end{array}$

}

\subsection{Qutrits $p=3$}
\flushleft{
$\begin{array}{|cc|cc|}
  \hline
  \multicolumn{4}{|l|}{\hspace{0.4cm}n=2}\\
  \hline
  d_1 & d_2 & c_1 & c_0 \\
  \hline
 2 & 0 & 1 & 2\\ 1 & 0 & 2 & 2\\
  \hline
\end{array}$ \\
\vspace{0.1cm}

$\begin{array}{|ccc|ccc|}
  \hline
  \multicolumn{6}{|l|}{\hspace{0.4cm}n=3}\\
  \hline
  d_1 & d_2 & d_3 & c_2 & c_1 & c_0\\
  \hline
  1 & 1 & 0 & 1 & 2 & 1\\ 2 & 1 & 1 & 2 & 0 & 1\\ 1 & 0 & 0 & 2 & 1 & 1 \\
  \hline
\end{array}$ \\
\vspace{0.1cm}

$\begin{array}{|cccc|cccc|}
\hline
  \multicolumn{8}{|l|}{\hspace{0.4cm}n=4}\\
  \hline
  d_1 & d_2 & d_3& d_4  &  c_3 & c_2 & c_1 & c_0 \\
  \hline
  1 & 1 & 0 & 1 & 0 & 0 & 1 & 2\\ 2 & 2 & 0 & 2 & 0 & 0 & 2 & 2\\ 1 & 2 & 1 & 1 & 1 & 0 & 0 & 2\\ 1 & 2 & 2 & 0 & 1 & 2 & 2 & 2 \\
  \hline
\end{array}$ \\
\vspace{0.1cm}

$\begin{array}{|ccccc|ccccc|}
\hline
  \multicolumn{10}{|l|}{\hspace{0.4cm}n=5}\\
  \hline
  d_1 & d_2 & d_3& d_4 & d_5  & c_4 & c_3 & c_2 & c_1 & c_0\\
  \hline
  2 & 1 & 2 & 0 & 1 & 0 & 0 & 0 & 2 & 1\\ 2 & 2 & 1 & 1 & 0 & 0 & 0 & 2 & 1 & 1\\ 0 & 1 & 2 & 0 & 0 & 0 & 1 & 0 & 1 & 1\\ 2 & 1 & 1 & 1 & 1 & 0 & 1 & 2 & 0 & 1 \\
  \hline
\end{array}$ \\
\vspace{0.1cm}

$\begin{array}{|cccccc|cccccc|}
\hline
  \multicolumn{12}{|l|}{\hspace{0.4cm}n=6}\\
  \hline
  d_1 & d_2 & d_3& d_4 & d_5 & d_6 & c_5 & c_4 & c_3 & c_2 & c_1 & c_0\\
  \hline
  1 & 0 & 2 & 2 & 1 & 0 & 0 & 2 & 1 & 1 & 1 & 2\\ 2 & 0 & 1 & 1 & 2 & 0 & 0 & 2 & 2 & 1 & 2 & 2\\ 1 & 0 & 2 & 0 & 2 & 0 & 1 & 0 & 0 & 0 & 0 & 2\\ 2 & 2 & 0 & 1 & 0 & 0 & 1 & 0 & 1 & 0 & 0 & 2 \\
  \hline
\end{array}$
}

\subsection{Qupits $p=5$}
\flushleft{
$\begin{array}{|cc|cc|}
  \hline
  \multicolumn{4}{|l|}{\hspace{0.4cm}n=2}\\
  \hline
  d_1 & d_2 & c_1 & c_0 \\
  \hline
 3 & 1 & 1 & 2\\ 4 & 2 & 4 & 2 \\
  \hline
\end{array}$ \\
\vspace{0.1cm}

$\begin{array}{|ccc|ccc|}
  \hline
  \multicolumn{6}{|l|}{\hspace{0.4cm}n=3}\\
  \hline
  d_1 & d_2 & d_3 & c_2 & c_1 & c_0\\
  \hline
 2 & 3 & 0 & 0 & 4 & 2\\ 3 & 2 & 0 & 0 & 4 & 3\\ 3 & 1 & 0 & 1 & 1 & 3\\ 4 & 2 & 3 & 1 & 4 & 3 \\
  \hline
\end{array}$ \\
\vspace{0.1cm}

$\begin{array}{|cccc|cccc|}
\hline
  \multicolumn{8}{|l|}{\hspace{0.4cm}n=4}\\
  \hline
  d_1 & d_2 & d_3& d_4  &  c_3 & c_2 & c_1 & c_0 \\
  \hline
 3 & 0 & 1 & 1 & 0 & 4 & 1 & 2\\ 1 & 3 & 0 & 1 & 0 & 4 & 4 & 2\\ 3 & 1 & 0 & 0 & 1 & 0 & 2 & 3\\ 2 & 3 & 2 & 1 & 2 & 0 & 3 & 3 \\
  \hline
\end{array}$ \\
\vspace{0.1cm}

$\begin{array}{|ccccc|ccccc|}
\hline
  \multicolumn{10}{|l|}{\hspace{0.4cm}n=5}\\
  \hline
  d_1 & d_2 & d_3& d_4 & d_5  & c_4 & c_3 & c_2 & c_1 & c_0\\
  \hline
  2 & 3 & 0 & 0 & 0 & 0 & 2 & 2 & 1 & 3\\ 3 & 2 & 0 & 0 & 0 & 0 & 2 & 3 & 1 & 2\\ 3 & 2 & 3 & 0 & 2 & 0 & 3 & 0 & 0 & 2\\ 3 & 0 & 2 & 3 & 2 & 0 & 3 & 0 & 0 & 3\\
  \hline
\end{array}$ \\
\vspace{0.1cm}

$\begin{array}{|cccccc|cccccc|}
\hline
  \multicolumn{12}{|l|}{\hspace{0.4cm}n=6}\\
  \hline
  d_1 & d_2 & d_3& d_4 & d_5 & d_6 & c_5 & c_4 & c_3 & c_2 & c_1 & c_0\\
  \hline
  4 & 2 & 2 & 4 & 1 & 2 & 0 & 0 & 0 & 1 & 1 & 3\\ 3 & 4 & 1 & 3 & 3 & 1 & 0 & 0 & 0 & 1 & 4 & 3\\ 1 & 3 & 2 & 0 & 4 & 0 & 0 & 0 & 1 & 2 & 0 & 2\\ 3 & 3 & 3 & 0 & 3 & 3 & 0 & 0 & 1 & 2 & 4 & 3\\
  \hline
\end{array}$
}

\subsection{Qupits $p=7$}
\flushleft{
$\begin{array}{|cc|cc|}
  \hline
  \multicolumn{4}{|l|}{\hspace{0.4cm}n=2}\\
  \hline
  d_1 & d_2 & c_1 & c_0 \\
  \hline
 4 & 1 & 2 & 3\\ 3 & 2 & 2 & 5\\ 6 & 3 & 5 & 3\\ 5 & 4 & 5 & 5\\
  \hline
\end{array}$ \\
\vspace{0.1cm}

$\begin{array}{|ccc|ccc|}
  \hline
  \multicolumn{6}{|l|}{\hspace{0.4cm}n=3}\\
  \hline
  d_1 & d_2 & d_3 & c_2 & c_1 & c_0\\
  \hline
 2 & 4 & 1 & 0 & 5 & 2\\ 3 & 3 & 1 & 0 & 6 & 2\\ 2 & 3 & 1 & 1 & 2 & 4\\ 6 & 3 & 3 & 2 & 1 & 4 \\
  \hline
\end{array}$ \\
\vspace{0.1cm}

$\begin{array}{|cccc|cccc|}
\hline
  \multicolumn{8}{|l|}{\hspace{0.4cm}n=4}\\
  \hline
  d_1 & d_2 & d_3& d_4  &  c_3 & c_2 & c_1 & c_0 \\
  \hline
5 & 4 & 4 & 1 & 0 & 3 & 3 & 3\\ 6 & 3 & 3 & 2 & 0 & 3 & 4 & 3\\ 6 & 0 & 5 & 3 & 0 & 4 & 3 & 3\\ 4 & 2 & 0 & 1 & 0 & 4 & 4 & 3 \\
  \hline
\end{array}$ \\
\vspace{0.1cm}

$\begin{array}{|ccccc|ccccc|}
\hline
  \multicolumn{10}{|l|}{\hspace{0.4cm}n=5}\\
  \hline
  d_1 & d_2 & d_3& d_4 & d_5  & c_4 & c_3 & c_2 & c_1 & c_0\\
  \hline
   5 & 1 & 0 & 1 & 0 & 0 & 0 & 0 & 2 & 2\\ 6 & 3 & 4 & 0 & 1 & 0 & 0 & 0 & 5 & 2\\ 6 & 4 & 0 & 1 & 3 & 0 & 0 & 2 & 2 & 4\\ 6 & 5 & 4 & 4 & 2 & 0 & 0 & 3 & 0 & 2 \\
  \hline
\end{array}$ \\
\vspace{0.1cm}

$\begin{array}{|cccccc|cccccc|}
\hline
  \multicolumn{12}{|l|}{\hspace{0.4cm}n=6}\\
  \hline
  d_1 & d_2 & d_3& d_4 & d_5 & d_6 & c_5 & c_4 & c_3 & c_2 & c_1 & c_0\\
  \hline
  6 & 6 & 0 & 1 & 0 & 1 & 0 & 0 & 0 & 3 & 1 & 5\\ 2 & 4 & 5 & 5 & 5 & 0 & 0 & 0 & 0 & 3 & 3 & 3\\ 5 & 3 & 2 & 2 & 2 & 0 & 0 & 0 & 0 & 3 & 4 & 3\\ 6 & 0 & 6 & 0 & 1 & 1 & 0 & 0 & 0 & 3 & 6 & 5 \\
  \hline
\end{array}$
}

\section{Example of MUBs for $d=3^3=27$ via symmetrized companion matrix} \label{lowdimensions1}
\begin{figure}[h]
\includegraphics[width=8cm]{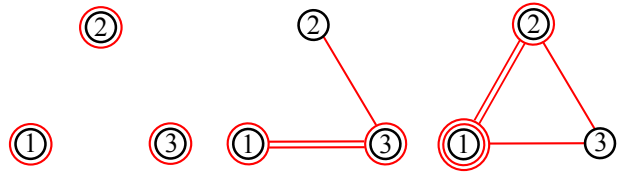}
\caption{(Color online) Fundamental graph--states of a complete set of MUBs for three qutrits derived in Appendix~\ref{lowdimensions1}. A complete set is obtained through all possible linear combinations over $\mathbb{Z}_3$.}
\label{3qutritsfund}
\end{figure}

We demonstrate the construction of a complete set of MUBs by the example of tripartite qutrit system, i.e. a Hilbert space of dimension $d=p^n$ where $p=3$ and $n=3$. According to the table in Appendix~\ref{listprimitive}, the polynomial $f(x)=x^3+x^2+2x+1$, having the coefficients $c_2=1$, $c_1=2$ and $c_0=1$, is irreducible over $\mathbb{Z}_3$. The corresponding companion matrix is
\begin{align}
C=\left(
    \begin{array}{ccc}
      0 & 1 & 0 \\
      0 & 0 & 1 \\
      -c_0 & -c_1 & -c_2 \\
    \end{array}
  \right)=\left(
    \begin{array}{ccc}
      0 & 1 & 0 \\
      0 & 0 & 1 \\
      2 & 1 & 2 \\
    \end{array}
  \right) \ .
\end{align}
Furthermore, the matrix $B_0$ as defined in Eq.~(\ref{B0matrix}) becomes
\begin{align}
B_0=\left(
\begin{array}{ccc}
 0 & 0 & 1 \\
 0 & 1 & 2 \\
 1 & 2 & 2 \\
\end{array}
\right) \ .
\end{align}
Since $\det(B_0)=2$ is a quadratic non--residue in $\mathbb{Z}_3$, and $(n \ \mathrm{mod} \ 4) =3$, we can choose $g=2$ for which the determinant of $B=g B_0$ becomes a quadratic residue. Applying the elimination procedure from Appendix~\ref{congruencep3} to the matrix $B$ to achieve $P B P^T= \mathbbm{1}_3$, one finds the matrix
\begin{align}
P=\left(
\begin{array}{ccc}
 0 & 0 & 1 \\
 0 & 1 & 2 \\
 1 & 2 & 2 \\
\end{array}
\right) \ ,  \hspace{0.5cm} P^{-1}=\left(
\begin{array}{ccc}
 2 & 1 & 1 \\
 1 & 1 & 0 \\
 1 & 0 & 0 \\
\end{array}
\right) \ .
\end{align}
Hence, a symmetric matrix $Q$ which is similar to the companion matrix $C$ is given by
\begin{align}
Q=P C P^{-1}= \left(
\begin{array}{ccc}
 1 & 0 & 2 \\
 0 & 0 & 1 \\
 2 & 1 & 1 \\
\end{array}
\right) \ .
\end{align}
Therefore, a basis in the symmetric matrix representation of the finite field $\F_{3^3}$ is given by the matrices $\{Q^0, Q^1, Q^2\}$, which are of the form
\begin{align}
Q^0= \left(
       \begin{array}{ccc}
         1 & 0 & 0 \\
         0 & 1 & 0 \\
         0 & 0 & 1 \\
       \end{array}
     \right) , \
Q^1= \left(
\begin{array}{ccc}
 1 & 0 & 2 \\
 0 & 0 & 1 \\
 2 & 1 & 1 \\
\end{array}
\right), \
Q^2= \left(
\begin{array}{ccc}
 2 & 2 & 1 \\
 2 & 1 & 1 \\
 1 & 1 & 0 \\
\end{array}
\right).
\end{align}
Thus, the $3$ fundamental adjacency matrices are $A_0=Q^0=\mathbbm{1}_3$, $A_1=Q^1$ and $A_2=Q^2$. These are illustrated in Fig.~\ref{3qutritsfund}. Consequently, a complete set of graphs is given by the $27$ different adjacency matrices $A_r$ from the set
\begin{align}
S= \left\{a_2 A_2 + a_1 A_1 + a_0 A_0\right\}_{a_0,a_1,a_2 \in \mathbb{Z}_3} \ .
\end{align}
Since the used polynomial $f(x)$ is also primitive, this set may equivalently be obtained via $S= \{ Q^i \}_{i=0}^{25} \cup \{ \mathds{O}_3 \}$. Subsequently, the bases $\{ \ket{G_r({m_1,m_2,m_3})} \}$ with the elements
\begin{align}
\ket{G_r({m_1,m_2,m_3})}= Z^{m_1} \otimes Z^{m_2} \otimes Z^{m_3} \ket{G_r} \ ,
\end{align}
where $m_1,m_2,m_3 \in \mathbb{Z}_3$, with
\begin{align}
\ket{G_r}= \prod_{i \leq j} U_{i,j}^{(A_r)_{i,j}} \ket{+}^{\otimes n} \ ,
\end{align}
defined by the adjacency matrix $A_r \in S$, are mutually unbiased. Together with the computational basis $\mathcal{B}_C= \{ \ket{0}, \ket{1}, \ket{2} \}^{\otimes 3}$, these bases form a complete set of $28$ MUBs.

\section{Example of MUBs for $d=2^3=8$ via a tridiagonal matrix} \label{lowdimensions2}
The construction of a complete set of MUBs is illustrated by the example of a tripartite qubit system, i.e. the Hilbert space of dimension $d=p^n$ with $p=2$ and $n=3$. Here, let us use the vector $\vec{d}=(1,0,0)$, from the table in Appendix~\ref{listprimitive}, for which the tridiagonal matrix from Eq.~(\ref{tridiag}) becomes
\begin{align}
Q=\left(
  \begin{array}{ccc}
    1 & 1 & 0 \\
    1 & 0 & 1 \\
    0 & 1 & 0 \\
  \end{array}
\right) \ .
\end{align}
The characteristic polynomial of this matrix is $f(x)=x^3+x^2+x+1$, which is irreducible (in addition, also primitive) over $\mathbb{Z}_2$. Therefore, we obtain the matrices
\begin{align}
Q^0= \left(
       \begin{array}{ccc}
         1 & 0 & 0 \\
         0 & 1 & 0 \\
         0 & 0 & 1 \\
       \end{array}
     \right),
Q^1= \left(
\begin{array}{ccc}
 1 & 1 & 0 \\
 1 & 0 & 1 \\
 0 & 1 & 0 \\
\end{array}
\right), \
Q^2= \left(
\begin{array}{ccc}
 0 & 1 & 1 \\
 1 & 0 & 0 \\
 1 & 0 & 1 \\
\end{array}
\right).
\end{align}
as a basis of the symmetric matrix representation of the finite field $\F_{2^3}$. Therefore, the matrix powers $0,1,2$ of $Q$ are the fundamental adjacency matrices, i.e. $A_0=Q^0$, $A_1=Q^1$ and $A_2=Q^2$. Thus, a complete set of graphs is given by the $8$ different adjacency matrices $A_r$ from the set
\begin{align}
S= \left\{a_2 A_2 + a_1 A_1 + a_0 A_0\right\}_{a_0,a_1,a_2 \in \mathbb{Z}_2} \ .
\end{align}
As the utilized polynomial $f(x)$ is also primitive, this set can equivalently be obtained via $S= \{ Q^i \}_{i=0}^{6} \cup \{ \mathds{O}_3 \}$. Now, the $8$ bases $\{ \ket{G_r({m_1,m_2,m_3})} \}$ with the elements
\begin{align}
\ket{G_r({m_1,m_2,m_3})}= Z^{m_1} \otimes Z^{m_2} \otimes Z^{m_3} \ket{G_r} \ ,
\end{align}
where $m_1,m_2,m_3 \in \mathbb{Z}_2$, and
\begin{align}
\ket{G_r}= \prod_{i \leq j} U_{i,j}^{\left(A_r\right)_{i,j}} \ket{+}^{\otimes n} \ .
\end{align}
defined via the adjacency matrices $A_r \in S$, are mutually unbiased. This set of bases is illustrated in Fig.~\ref{3qubitsgraph}. Together with the computational basis $\mathcal{B}_C=\{\ket{0}, \ket{1} \}^{\otimes 3}$, we have a complete set of $9$ MUBs.

\end{document}